\documentclass[twocolumn,showpacs,preprintnumbers,amsmath,amssymb]{revtex4-1}

\usepackage{float}
\usepackage{graphicx}
\usepackage{dcolumn}
\usepackage{bm}
\usepackage{color}
\usepackage{xcolor}

\usepackage{amsmath}
\usepackage{amssymb}
\usepackage{amsfonts}
\usepackage{bbm}
\usepackage{hyperref}

\usepackage{booktabs}

\begin{document}

\title{A laser with  instability reaching $4 \times 10^{-17}$ based on a 10-cm-long silicon cavity at sub-5-K temperatures}

\author{Zhi-Ang Chen$^{1,2}$}
\author{Hao-Ran Zeng$^{1,2}$}
\author{Wen-Wei Wang$^{1,2}$}
\author{Han Zhang$^{1,2}\dagger$}
\author{Run-Qi Lei$^{1,2}$}
\author{Jian-Zhang Li$^{3}$}
\author{Cai-Yin Pang$^{3}$}
\author{She-Song Huang$^{4}$}
\author{Xibo Zhang$^{1,2,5,6}$}
\email{Corresponding author. Email: xibo@pku.edu.cn\\$\dagger$Present address: Quantum Science Center of Guangdong-Hong Kong-Macao Greater Bay Area, Shenzhen 518045, China (H.Z.).} 
\affiliation{$^1$International Center for Quantum Materials, School of Physics, Peking University, Beijing 100871, China}
\affiliation{$^2$Collaborative Innovation Center of Quantum Matter, Beijing 100871, China}

\affiliation{$^3$Beijing Xinnan Zhike Optoelectronic Technology Co., Ltd, Beijing 100023, China}
\affiliation{$^4$Beijing Physike Technology Co., Ltd, Beijing 100085, China}
\affiliation{$^5$Hefei National Laboratory, Hefei 230088, China}
\affiliation{$^6$Beijing Academy of Quantum Information Sciences, Beijing 100193, China}

\begin{abstract}
The realization of ultra-stable lasers with $10^{-17}$-level frequency stability has enabled a wide range of researches on precision metrology and fundamental science, where cryogenic single-crystalline cavities constitute the heart of such ultra-stable lasers. For further improvements in stability, increasing the cavity length at few-kelvin temperatures provides a promising alternative to utilizing relatively short cavities with novel coating, but has yet to be demonstrated with state-of-the-art stability. Here we report on the realization of a relatively long ultra-stable silicon cavity with a length of 10~cm and sub-5-K operating temperatures. We devise a dynamical protocol of cool-quiet quench measurement that reveals the inherent $10^{-17}$-level frequency instability of the silicon cavity despite the substantially larger frequency noise induced by the cryostat vibration. We further develop a method for suppressing the cryostat-vibration-induced frequency noise under continuous cooling, and observe an average frequency instability of $4.3(2) \times 10^{-17}$ for averaging times of 4 to 12~seconds. Using the measured noise power spectral density, we compute a median linewidth of 9.6(3)~mHz for the silicon cavity laser at 1397~nm, which is supported by an empirically determined linewidth of 5.7(3)~mHz based on direct optical beat measurements. These results establish a new record for optical cavities within a closed-cycle cryocooler at sub-5-K temperatures and provide a prototypical system for using long cryogenic cavities to enhance frequency stabilities to the low-$10^{-17}$  or better level.

\end{abstract}


\maketitle

\noindent\textbf{Article history:}\\
\noindent{Received: 05-Jun-2025}\\
\noindent{Revised: 21-Jul-2025}\\
\noindent{Accepted: 13-Aug-2025}\\

\\

\noindent\textbf{1. Introduction}\\

Advances in laser frequency stabilization have led to rapid progress in the studies of optical atomic clocks~\cite{Ludlow15rmp,Zhang16nsr}, precision metrology~\cite{Hall06rmp,Hansch06rmp}, 
and fundamental science~\cite{Blatt08prl, Chou10science,McGrew18nature,Bothwell22nature}. In particular, highly stable optical local oscillators (namely ultra-stable lasers) are an indispensable ingredient of optical clock experiments that demonstrate excellent measurement precision with statistical frequency uncertainties on the order of  $10^{-19}$ or better~\cite{McGrew18nature,Oelker19nph,Bothwell22nature,Zheng22nature}. Recently, state-of-the-art ultra-stable lasers at wavelengths near 1542~nm have reached frequency stabilities in the middle of the $10^{-17}$ range, which are fundamentally limited by the Brownian thermal noise of cryogenic single-crystalline silicon optical cavities with dielectric coatings~\cite{Matei17prl, Zhang17prl, Robinson19optica}. Cryogenic silicon cavities with semiconductor crystalline coatings~\cite{Cole13nph} hold promise for realizing an even lower level of Brownian thermal noise. However, reaching such a low thermal noise limit in the associated laser frequency stability has been encumbered by new noise sources like birefringent noise and other unclear residual noise~\cite{Kedar23optica, Yu23prx}. Several alternative methods have also been studied, including engineering the cavity temperature~\cite{Robinson21ol, Gillot23jointconf}, choosing different crystalline materials~\cite{He23ol, Valencia24rsi} and exploring new coating structures~\cite{Dickmann23commphys}.

With a characteristic $\sqrt{T}/L$ dependence on the cavity length $L$ and cavity temperature $T$, the fundamental Brownian thermal noise can be suppressed by increasing $L$ and reducing $T$. This is straightforward to consider, but very challenging to realize at liquid-helium temperatures, which is due to the difficulty in simultaneously achieving a relatively long cavity and effectively mitigating the increased frequency noise induced by the corresponding large-volume cryostat.
Compared with the apparatus in Refs.~\cite{Zhang17prl, Robinson19optica, Kedar23optica}, a new cryostat was recently designed and realized that reaches lower operating temperatures of about 3~K for a substantially longer silicon cavity with a length of 10~cm~\cite{Wang23fip}. Progress on 1.7-K and sub-kelvin cryogenic silicon cavities has also been reported~\cite{Wiens23oe, Barbarat24arxiv}. However, $10^{-17}$-level laser frequency stability has yet to be demonstrated with relatively long cavities at few-kelvin temperatures.

\begin{figure*}[t]
	\includegraphics[width = 17.2cm]{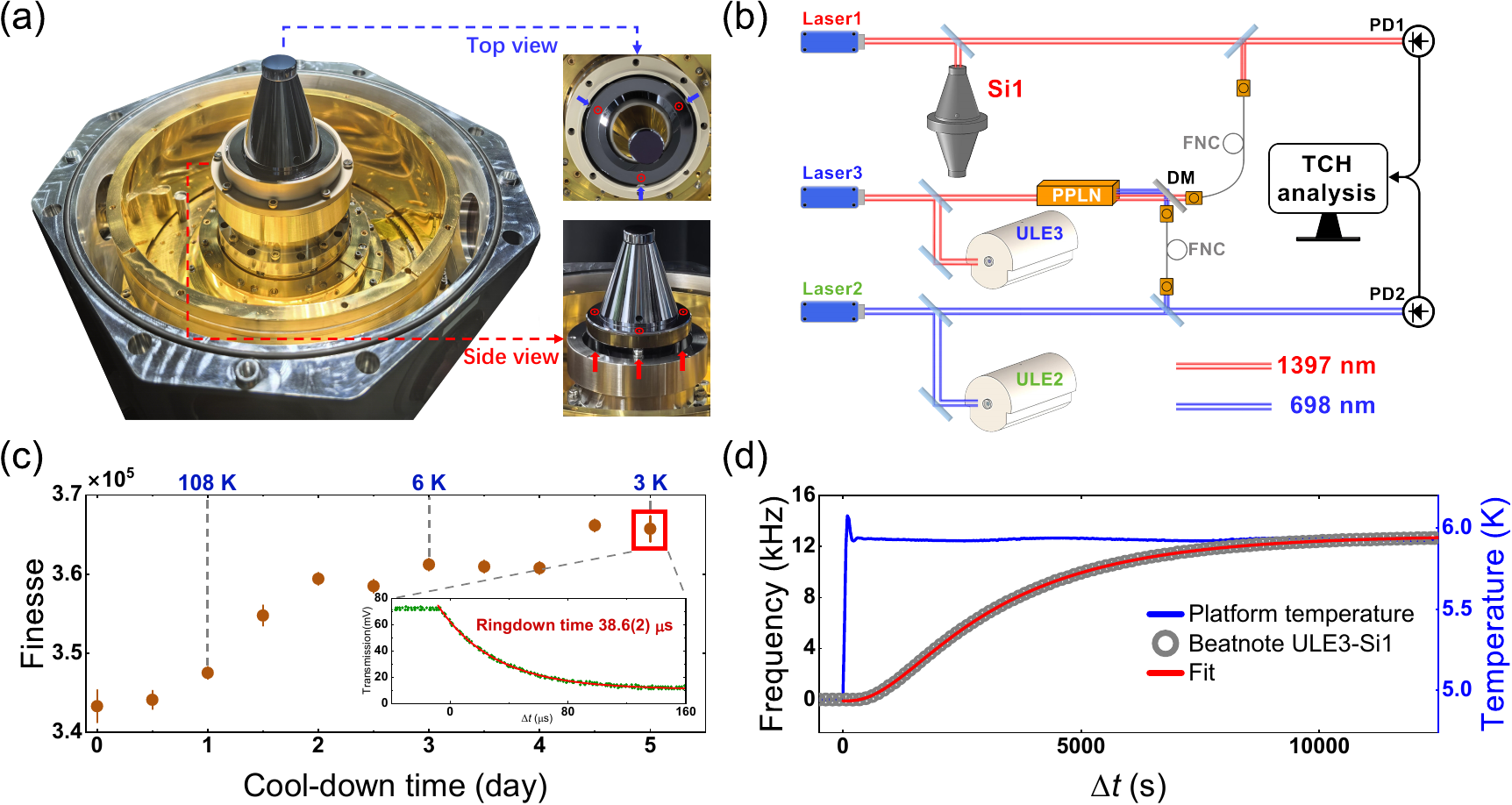}
	\caption{\label{fig:setup}  Experimental setup and safe cooling down of a silicon cavity to few-kelvin temperatures with intact optical performance. (a) Silicon cavity (Si1) and its vacuum chamber for cryogenic cooling. In the insets, red arrows indicate a vertical support under the Si1 central ring at three points marked by centers of red circles, and blue arrows indicate a three-point fastening of the Si1 horizontal position (not shown in the side view).  (b) Diagram of three-cornered-hat (TCH) frequency stability measurements based on the Si1 cavity and two room-temperature ultra-low-expansion (ULE) glass cavities, ULE2 and ULE3. DM: dichroic mirror; PD: photo-detector; FNC: fiber noise cancellation; PPLN: periodically poled lithium niobate waveguide for frequency doubling. (c) The Si1 cavity finesse (solid circles) remains stable and slightly increases during the cooling. Inset: a ringdown measurement for determining the finesse at a platform temperature of about 3~K. (d) Step-function response of the Si1 resonance frequency for \textit{in situ} determination of cavity temperature, with improvements (1) to (4) implemented. The blue line is the platform temperature. Gray circles denote the beat frequency between Laser3 (based on the ULE3 cavity) and Laser1 (Si1). The red line shows a fit using a third-order low-pass filter model. Error bars represent $1\sigma$ statistical uncertainties.
	}
\end{figure*}

In this article, we experimentally realize a 10-cm-long, sub-5-K silicon optical cavity with a median laser linewidth below 10~mHz at 1397~nm and a frequency instability of $4\times 10^{-17}$. We first develop a dynamical experimental protocol for performing three-cornered-hat (TCH) measurements and demonstrate a $10^{-17}$-level
optical performance for the cryogenic silicon cavity even when the cryostat-vibration-induced frequency noise is at a substantially higher level. Under continuous cooling, we further identify two major uncontrolled  noise sources that seriously hamper the stability of the silicon cavity, and accordingly demonstrate strategies that successfully overcome these  obstacles. The resultant silicon cavity shows an optimized frequency instability of $4.3(2)\times 10^{-17}$ at averaging times of 4 to 12 seconds, which is close to the Brownian thermal noise floor of this cavity at 4.9~K (about $3.3\times 10^{-17}$; see the Supplementary materials) and revealing better frequency stability at these averaging times than the previous best sub-5-K optical cavities~\cite{Zhang17prl,Robinson19optica,Kedar23optica}.
Our work offers a promising platform for reaching  laser stabilities in the low-$10^{-17}$ range or better based on relatively long silicon cavities at even lower temperatures. \\

\noindent\textbf{2. Setup and methods}\\

As shown in Fig.~\ref{fig:setup}a, we designed and realized a 10-cm-long ultra-stable silicon optical cavity, referred to as Si1.
The monocrystalline cavity spacer 
has an optical axis oriented along the crystalline $\langle 111 \rangle$ axis of silicon. A pair of high-reflectivity cavity mirrors, with radii of curvature of $\infty$ and 1~m respectively, are superpolished,
coated with a dielectric coating technique,
 and subsequently optically contacted to the 9.79-cm-long spacer (calibrated under about 3~K), providing a finesse of about $3.6\times 10^5$. This cavity operates at sub-5-K temperatures in a two-stage
 Gifford-McMahon closed-cycle cryostat~\cite{Wang23fip}.

To prepare a clean and low-vibration few-kelvin cryogenic platform to accommodate the Si1 cavity, we implemented six technical improvements, denoted as improvements (1) to (6), with respect to our previous cryogenic apparatus~\cite{Wang23fip} and an earlier version of a cryogenic silicon cavity (named Si0): 

(1) To eliminate harmful outgassing processes that potentially contaminate the cavity mirrors, we thoroughly cleaned all in-vacuum surfaces, abandoned certain ``standard practices'' of the cryostat manufacturer Montana Instruments, and applied only high-vacuum-compatible material to maintain good thermal contact between adjacent surfaces (see the Supplementary materials).

(2) To further suppress the condensation of gas molecules onto the cavity mirrors, we adopted a stringent vacuum criterion by reducing the in-sample-chamber pressure to the $10^{-7}$-Torr high-vacuum regime before any cooling started (see the Supplementary materials).

(3) To improve the rigidity of the mechanical structure supporting the silicon cavity vertically (red arrows in Fig.~\ref{fig:setup}a inset), we built a semi-monolithic, all-stainless-steel, new supporting structure that has no shakable components within itself and  realizes higher mechanical resonance frequencies (see the Supplementary materials).

(4) To reduce the vibration transferred from cryostat coolers, we separated the baseplate of the sample chamber from that of the cryocooler chamber, and enabled an active vibration control
platform (Table Stable, AVI-200-XL-LP), on which the sample chamber sat.

(5) To further suppress the frequency noise induced by cryostat cooler vibration, we designed and implemented a three-point fastening of the horizontal position of the Si1 cavity (Fig.~\ref{fig:setup}a, upper inset), with proper strain relief applied (see the Supplementary materials). 

(6) To protect the cavity against  environmental fluctuations such as acoustic noise and air flow, we enclosed the entire optical path of the silicon cavity laser inside an acoustic shield made of sound-absorbing cotton.

Table~\ref{table:SiCavityConfig} lists a few configurations with various numbers of implemented improvements and experimental conditions.

We realized an ultra-stable laser (``LoongLaser-I'', or simply ``Laser1'') by phase-locking it to the Si1 cavity based on the Pound-Drever-Hall (PDH) approach~\cite{DH83apb}. 
This laser (operated near 1396.8892~nm) was locked to the TEM${}_{00}$ mode of Si1 with a 600-kHz locking bandwidth and controlled $\mu$W-level incident optical power (see the Supplementary materials).
To measure frequency stabilities, we built two other stable lasers (``Laser2'' and ``Laser3'') that were independently phase-stabilized to two room-temperature, 30-cm ultra-low-expansion (ULE) glass cavities, referred to as ULE2~(operated near 698.4445~nm, close to the strontium clock transition) and ULE3~(near 1396.8900~nm) respectively;
see Fig.~\ref{fig:setup}b. The three-laser system enabled us to measure optical beat frequencies (using K+K FXE frequency counters) and perform TCH frequency stability analysis 
 to determine the frequency instability for each laser, represented by the modified Allan deviation ~\cite{Premoli93IEEE,Zhang17prl}; see the Supplementary materials. Here, the frequencies of the 698-nm Laser2 and 1397-nm Laser3 were bridged by frequency doubling via a single-pass periodically poled lithium niobate (PPLN) waveguide~\cite{Zhu24rsi}. All  frequency synthesizers and counters were referenced to a 10-MHz rubidium clock (Chengdu Synchronization Technology, STM-Rb-N).\\

\begin{table}
	\center
	\caption{Experimental configurations of the silicon cavity}
	\label{table:SiCavityConfig}
	\begin{tabular}{llll} 
		\specialrule{1.0pt}{0pt}{0pt}
		Cavity & Implemented   & Cavity & Related figures \\
		configuration & improvements & temperature & \\
		\specialrule{0.5pt}{0pt}{0pt}
		Si1a &  (1) to (4) 
		& 4.9~K & Figs.~3, 1d  \\
		Si1b &   (1) to (5)  & 4.9~K & 
	Figs.~3, 4b \\
		Si1c &  (1) to (6) & 4.9~K & 
		Fig.~4  \\
		Si1c$'$ & (1) to (6) & 3.3~K  & Fig.~4a inset \\
		Si1-CQQM & (1) to (4) & near 3.3~K & Figs.~2, 3 \\
		Si0 & none & 3~K & Fig.~3  \\
        \specialrule{1.0pt}{0pt}{0pt}
	\end{tabular}
\end{table}

\noindent\textbf{3. Results}\\

\noindent\textit{3.1. Cooling the Si1 cavity without optical contamination}\\

While a silicon cavity operating at several kelvins naturally serves as a powerful cryogenic pump, such pumping can potentially lead to optical contamination of the cavity mirrors. To protect these mirrors, we prepared a clean environment by implementing improvements (1) and (2),
and calibrated the initial optical properties of the Si1 cavity under room temperature.
For the TEM${}_{00}$ mode, we performed ringdown measurements~\cite{Anderson84ao,Martin13thesis} under low and high vacuum with pressures of $P_{\mathrm{LV}}\approx 4$~Torr and $P_{\mathrm{HV}}\approx 4\times 10^{-7}$~Torr, and determined the effective finesses of the Si1 cavity to be $\mathcal{F}_{1,\mathrm{LV}}=1.68(10)\times 10^5$ and $\mathcal{F}_{1,\mathrm{HV}}=3.43(3)\times 10^5 > \mathcal{F}_{1,\mathrm{LV}}$ at these pressures respectively, which is qualitatively consistent with the influence of water absorption near 1397~nm. The Si1 cavity finesse for the TEM${}_{01}$ mode is similar to that for TEM${}_{00}$ with a fractional difference of $-3.5\%$, demonstrating good uniformity of mirror finesse. 

The cleanness of the high-vacuum environment was enhanced during a five-day cooling process, which was evidenced by the robustness of the Si1 optical properties.
During the cooling, the pressure decreased to about $1\times 10^{-8}$~Torr, and the coupling efficiency of Laser1 to the Si1 TEM${}_{00}$ mode remained unchanged within the measurement uncertainty without reoptimization.
As shown in Fig.~\ref{fig:setup}c, the measured Si1 cavity finesse slowly increased to a value $7\%$ larger than  $\mathcal{F}_{1,\mathrm{HV}}$ under a cryostat platform temperature of about 3~K.
Under this platform temperature, we further estimated a mean loss coefficient $L_{\mathrm{M}} \approx 3$~ppm and a mean transmission coefficient $T_{\mathrm{M}} \approx 5$~ppm~$> L_{\mathrm{M}}$ of the cavity mirrors for the TEM${}_{00}$ mode
	(see the Supplementary materials). 
Furthermore, even after one full thermal cycle from room temperature to several kelvins (with a four-month cryogenic operation period) and back to room temperature, the Si1 cavity finesse at high vacuum remained $\mathcal{F}'_{1,\mathrm{HV}} = 3.42(3) \times 10^5$, showing almost no sign of degradation with respect to $\mathcal{F}_{1,\mathrm{HV}}$. These observations support that 
our clean cryogenic environment has vastly suppressed potential optical contamination and kept the Si1 cavity optically intact at  temperatures of a few kelvins. In sharp contrast, the earlier Si0 cavity showed subtle evidences of contamination, exhibiting substantially lower and spatially inhomogeneous finesse that further degraded after a few months (see the Supplementary materials).

To quantitatively characterize the average \textit{in situ} temperature of the whole Si1 cavity, we identified the cryostat platform temperature as a convenient approximate ``readout port'' for this goal based on the following considerations.
First, we have previously measured the cavity temperature at its center ring and found it to be fairly close to the  platform temperature (0.13~K above the latter)~\cite{Wang23fip}.
Here, this sensor was removed to eliminate the associated thermal conduction path, and we expected the cavity to be more efficiently cooled to a slightly lower temperature that is even closer to the platform temperature. Second, we determined the average \textit{in situ} Si1 cavity temperature using a quench technique based on measurements of cavity resonance frequencies and knowledge of the silicon coefficient of thermal expansion (CTE)~\cite{Kedar23thesis, Hagemann13thesis}. As shown in Fig.~\ref{fig:setup}d, with improvements (1) to (4) implemented, we suddenly changed the set platform temperature by 1~K from 4.94 to 5.94~K, and observed a smooth crossover of the Laser3-Laser1 beat frequency from one value to another. This crossover behavior can be fitted using a third-order low-pass filter~\cite{Wang23fip} with time constants of 500~s, 500~s, and 2700~s, and a total frequency change of 12.7~kHz can be used to determine the cavity temperature. Specifically, we used a model of cryogenic material properties by the National Institute of Standards and Technology
(
https://trc.nist.gov/cryogenics/calculators/propcalc.html/
), which is supported by experiments~\cite{Swenson83JPCRD,Wiens14ol,Wiens15ol}. This model shows that, for low temperatures ($T\le 10$~K), the silicon CTE can be approximately described by a $T^3$ scaling, namely, $\mathrm{CTE}_{\mathrm{Si}} \approx \alpha T^3$, where $\alpha$ is a coefficient. 
Integrating this scaling with respect to $T$ leads to a total frequency change of $\frac{\alpha}{4}[(T_{\mathrm{i}}+1)^4-T_{\mathrm{i}}^4]$, with $T_{\mathrm{i}}$ being the initial cavity temperature. Based on a  recent study on  ultra-stable silicon cavities at several kelvins~\cite{Kedar23thesis}, where $\mathrm{CTE}_{\mathrm{Si}}$ was measured to be $4.5\times 10^{-11}/\mathrm{K}$ at 4.9~K, we extracted $\alpha \approx 3.8\times 10^{-13}/\mathrm{K}^4$ and thus determined $T_{\mathrm{i}} \approx 4.9$~K for the Si1 cavity, which was indeed close to the initial platform temperature of 4.94~K. In another quench measurement, we determined $T_{\mathrm{i}}' \approx 3.1$~K, again close to the corresponding initial platform temperature of 3.28~K. Taking these into account, we chose the platform temperature as a convenient characterization of the Si1 cavity temperature $T$ in the sub-5-K regime.\\

\begin{figure}[t]
	\includegraphics[width = 8.6cm]{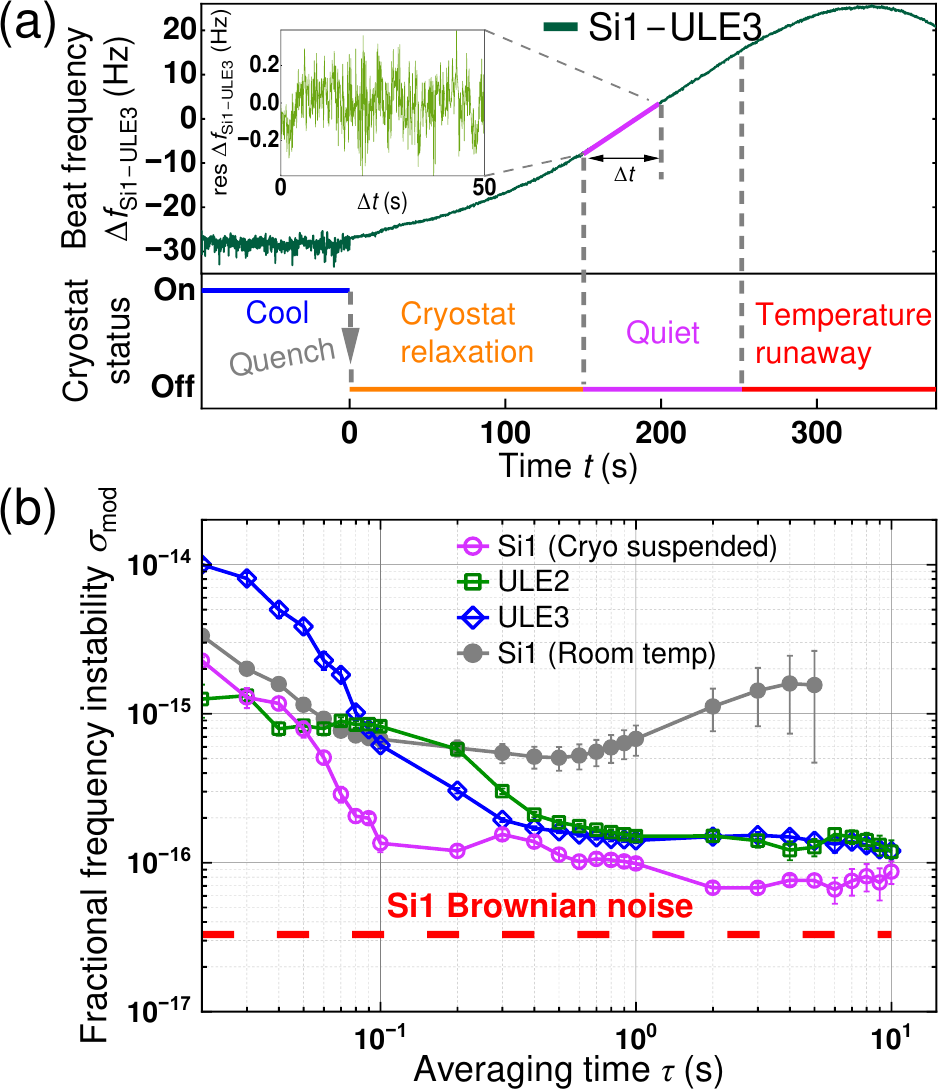}
	\caption{\label{fig:cqqm} Dynamically revealing the $10^{-17}$-level frequency stability of a cryostat-vibration-free Si1 cavity by cool-quiet quench measurement (CQQM).
 (a) Diagram of the CQQM. Here, $\Delta f_{\mathrm{Si1-ULE3}}$ denotes an optical beat frequency between Laser1 and Laser3 (mixed down to near d.c.), and res~$\Delta f_{\mathrm{Si1-ULE3}}$ in the inset further denotes the residual frequency noise after subtracting an  offset and an overall linear drift from the $\Delta f_{\mathrm{Si1-ULE3}}$ data. (b) Frequency stabilities of the Si1, ULE2, and ULE3 cavities. The vertical axis corresponds to the modified Allan deviation. A dashed line shows a typical Brownian thermal noise floor of the Si1 cavity at 4.9~K. Magenta empty circles denote the frequency instability of the Si1 cavity measured by CQQM  (illustrated by the 50-second-long data set in the inset of (a)), which bypasses the cryostat vibration and reveals a sub-total effect mainly contributed by the optical performance of the Si1 cavity and  electro-optically induced technical noise sources. Here, improvements (1) to (4) are implemented, while (5) and (6) are not (see Table~\ref{table:SiCavityConfig}). Olive squares and blue diamonds denote the corresponding frequency stabilities of the ULE2 and ULE3 cavities respectively. For comparison, gray solid circles show the substantially larger frequency instability of the Si1 cavity under a room temperature of 292~K. Error bars represent $1\sigma$ statistical uncertainties.
	}
\end{figure}

\noindent\textit{3.2. Dynamically revealing the $10^{-17}$-level frequency stability of a cryostat-vibration-free Si1 cavity by  ``Cool-Quiet'' quench measurement}\\

At sub-5-K temperatures, the Si1 cavity has a fundamental Brownian thermal noise floor at and below $3.3\times 10^{-17}$, which is computed using $T = 4.9$~K, $L = 9.79$~cm, a 1397-nm wavelength and dielectric coating properties similar to those in Ref.~[\citenum{Robinson21ol}]; see Fig.~\ref{fig:cqqm} and the Supplementary materials. Under these conditions, the Si1 cavity is estimated to have a frequency drift rate below $1$~mHz/s; see the Supplementary materials. Because vibrational noise due to the cryostat can often overwhelm such a low Brownian thermal noise floor~\cite{Wiens23oe}, it is desirable to develop a diagnostic method for demonstrating and verifying the high optical performance of such cavities without being restricted by cryostat vibration.

%
We developed a protocol of cool-quiet quench measurement (CQQM) to determine, in a manner insensitive to the cryostat vibration, the Laser1 frequency stability due to the fundamental optical performance of the Si1 cavity and precision of the associated electro-optical control system. In this protocol shown in Fig.~\ref{fig:cqqm}a, the Si1 cavity was initially cooled to a low temperature in the range of a few kelvins (namely the ``Cool'' stage), where the whole cryostat vibration effect was contained in the Laser1 frequency noise. Then the cryostat cooling was suddenly disabled at time $t = 0$ (namely the ``Quench'' action), after which the compressor and other parts took some time to relax into a new equilibrium. When the new, quieter mechanical equilibrium was reached, we performed TCH measurements to extract the Laser1 frequency instability with the cryostat  suspended and the corresponding vibration temporarily removed (namely the ``Quiet'' stage). Here, the measurement time was  properly constrained such that the cavity temperature did not drift too much.

\begin{figure*}[t]
	\includegraphics[width = 17.2cm]{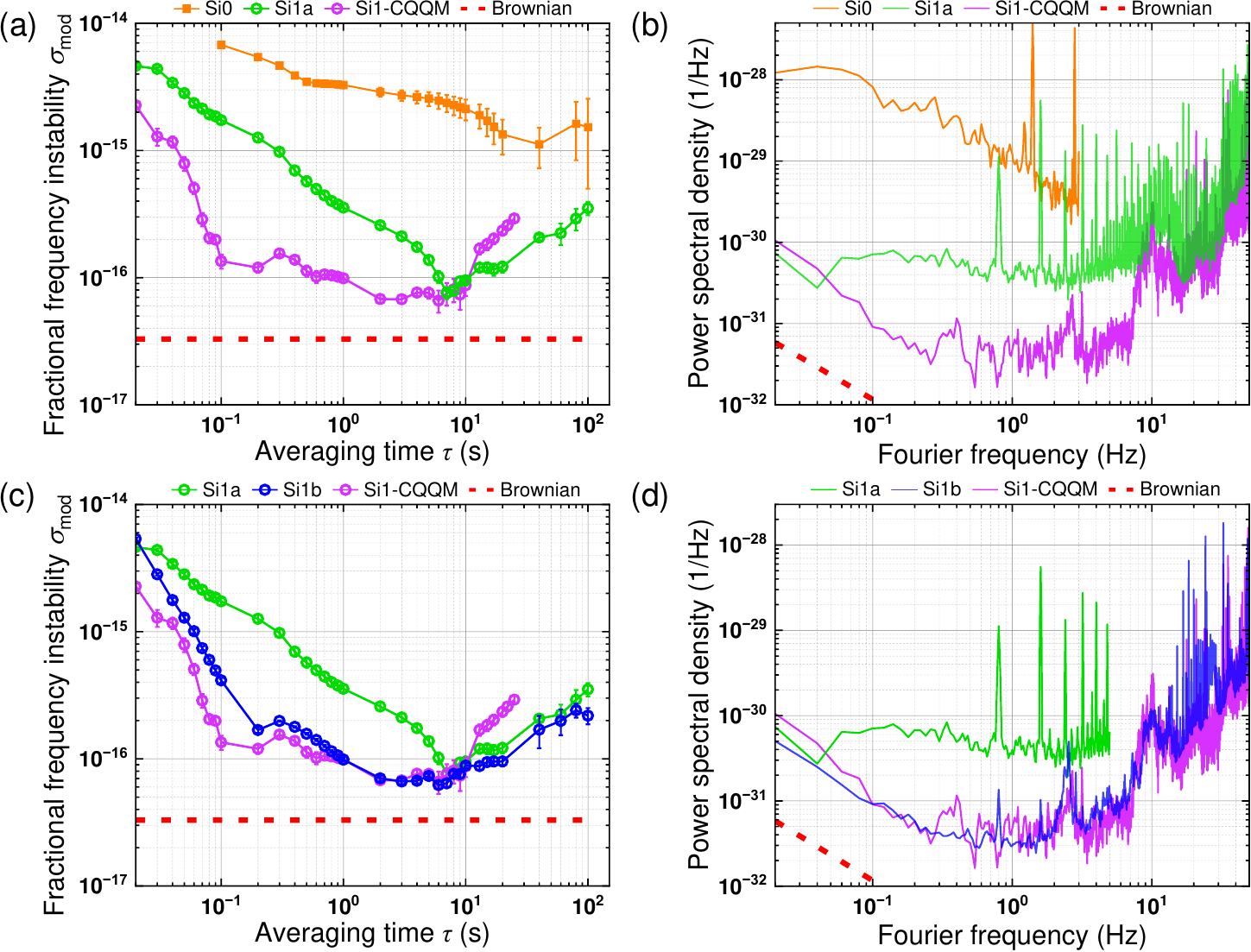}
	\caption{\label{fig:twomajorimprovements} Overcoming two major obstacles to realizing $10^{-17}$-level frequency stability under continuous cooling.		
(a) Modified Allan deviations. With technical improvements (1) to (4) implemented, the frequency stability of the silicon cavity (``Si1a'' configuration, green empty circles) has significantly improved with respect to that of the earlier Si0 cavity with mirror contamination (orange solid squares).  (b) Power spectral densities (PSDs) for measurements in (a). (c) Modified Allan deviations. With technical improvements (1) to (5) implemented, the frequency stability of the silicon cavity (``Si1b'' configuration, blue empty circles) further shows  stability enhancement over Si1a (by a factor of more than three at $\tau = 1$~s). (d) PSDs for measurements in (c). To better show the Si1b PSD, the Si1a PSD is only plotted in part in (d). 
	In (a) to (d), the Brownian thermal noise floor  is shown as red dashed lines, and the CQQM results, as a reference for comparison, are shown as magenta empty circles (modified Allan deviations) and magenta lines (PSDs).  
 Error bars represent $1\sigma$ statistical uncertainties.
	}
\end{figure*}

We performed CQQM with an initial platform temperature of 3.3~K and observed $10^{-17}$-level frequency instability for the Si1 cavity with improvements (1) to (4) implemented. As shown in Fig.~\ref{fig:cqqm}b, the Si1 frequency instability reached $9.9(7)\times 10^{-17}$ at an averaging time of $\tau = 1$~s, improved to $6.8(5)\times 10^{-17}$ at 2~s and $6.6(1.3)\times 10^{-17}$ at 6~s, and changed towards $1\times 10^{-16}$ at $7 \sim 10$~s. Such stability not only surpassed Si1's  own room-temperature stability, but also outperformed the reference cavities ULE2 and ULE3, which both showed frequency instabilities below $2\times 10^{-16}$ at $1 \sim 10$~s. The Si1 instability revealed by CQQM reached the same order of magnitude as the typical Brownian thermal noise floor at 4.9~K (red dashed line in Fig.~\ref{fig:cqqm}b), which signifies that, except for the cryostat vibration, most noise sources have been well controlled to the $10^{-17}$ level. By contrast, 
 neither the $10^{-17}$-level CQQM result nor the $10^{-16}$-level instability at room temperature has ever been observed for the Si0 cavity. 
In next sections, we chose a platform temperature of  4.9~K under a ``slow'' cooling mode with minimum cryostat vibration. 
\\

\noindent\textit{3.3. Overcoming two obstacles to achieving $10^{-17}$-level frequency stability under continuous cooling}\\

With continuous cooling of the cryostat,
we identified two major obstacles as two types of uncontrolled technical perturbations that are difficult to model but severely degrade the frequency stability of a silicon cavity at a few kelvins. 
The first obstacle lies in 
uncontrolled optical scattering due to mirror contamination.
The second stems from uncontrolled vibration
amplified by non-rigidity in the whole assembly of the cavity and its supporting structures.
 Below we depict two rounds of Si1 stability enhancement after addressing these obstacles through the implementations of improvements (1) to (5).

In the first round, we eliminated mirror contamination and improved the rigidity of a vertical supporting structure for the cavity through improvements (1) to (4). Accordingly, the frequency instability of the silicon cavity (denoted as ``Si1a'') reached $3.6(3)\times 10^{-16}$ at $\tau = 1$~s and $7.6(8) \times 10^{-17}$ at around 7~s, representing a significant enhancement with respect to the $10^{-15}$-level Si0 instability;
see Fig.~\ref{fig:twomajorimprovements}a.
We further observed the distinct frequency noise power spectral densities (PSDs) of Si1a and Si0. As shown in Fig~\ref{fig:twomajorimprovements}b, Si0 exhibited an apparent flicker noise at Fourier frequencies of 0.1 to several hertz, 
which is difficult to model and primarily related to uncontrolled optical scattering with contaminated mirrors.
With new mirrors, Si1a showed a much lower, rather flat broadband noise background in its PSD over Fourier frequencies of 0.06 to 5~Hz, which is nevertheless
still larger than the reference PSD of Si1-CQQM. The Si1a-versus-Si1-CQQM differential measurements pointed to cryostat vibration as the major limiting factor for Si1a.

\begin{figure*}[t]
	\includegraphics[width = 17.2cm]{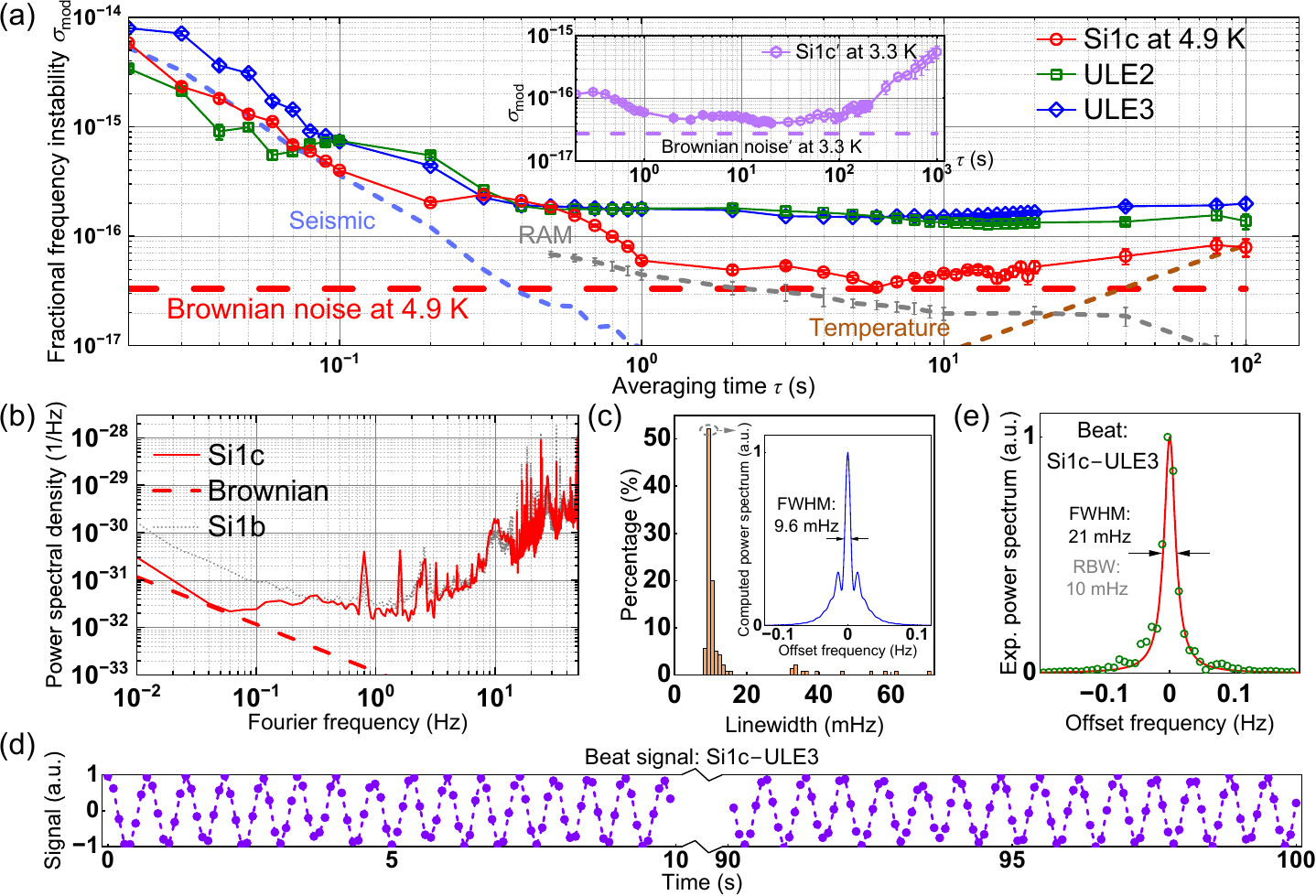}
	\caption{\label{fig:contiunous}  Optimized frequency instability approaching the Si1 thermal noise floor at 4.9~K under continuous cooling. (a) Modified Allan deviations. Measured frequency instabilities are shown for the silicon cavity with improvements (1) to (6) implemented (``Si1c'', red circles) as well as for ULE2 (olive squares) and ULE3 (blue diamonds) cavities. Red, blue, gray, and brown dashed lines show the computed Brownian thermal noise floor at 4.9~K (as also shown in Figs.~\ref{fig:cqqm} and \ref{fig:twomajorimprovements}), the estimated instability induced by seismic vibrations, the measured instability due to residual amplitude modulation (RAM), and the projected instability due to temperature fluctuation. The inset shows the Si1 frequency instability (purple hexagons) in a configuration (denoted as Si1c$'$) that is almost the same as Si1c except for a reduced platform temperature of 3.3~K.  (b) PSDs. The Si1c spectrum (solid line) is compared with that of Si1b (short dotted line) and the Brownian thermal noise floor.  (c) Based on the Si1c PSDs (measured for (b)) and a systematic scheme in Ref.~[\citenum{Bishof13prl}], the Si1c linewidths are computed and the histogram constructed, with a median full width at half-maximum (FWHM) of 9.6(3)~mHz (shown in the inset of (c)). (d) A sample 100-s-long beat signal (mixed down to near d.c. and measured with a digital oscilloscope) between the Si1 and ULE3 cavities. (e) Average line profile (green circles) for 12 data sets similar to the beat signal measured in (d), analyzed via normalized fast Fourier transform  using a 10-mHz resolution bandwidth (RBW) and Hanning window. The profile is fitted to a Lorentzian form (red line).
 Error bars represent $1\sigma$ statistical uncertainties.} 
\end{figure*}

In the second round, we further addressed the issue of cryostat vibration by sufficiently increasing the overall rigidity of the whole cavity-and-support assembly through improvement (5) (illustrated in Fig.~\ref{fig:setup}a). With improvements (1) to (5) all implemented, we observed an overall stability enhancement for the silicon cavity (denoted as Si1b) with respect to Si1a. The Si1b instability reached $9.8(3)\times 10^{-17}$ at $\tau = 1$~s and improved to $6.2(4)\times 10^{-17}$ at 6~s (Fig.~\ref{fig:twomajorimprovements}c).  Remarkably, at Fourier frequencies below 20~Hz, the Si1b  PSD reached a level very similar to that of the reference PSD of Si1-CQQM (Fig.~\ref{fig:twomajorimprovements}d).
From the differential PSD of Si1b versus Si1-CQQM, we estimated a vastly suppressed residual cryostat-vibration effect that corresponds to a modified Allan deviation below $2 \times 10^{-17}$ for $\tau \ge 3$~s (see the Supplementary materials). This second enhancement can be qualitatively understood as follows. For relatively long cryogenic cavities like Si1, in the absence of a horizontal supporting or constraining structure, the frequency stability 
becomes susceptible to cryostat-induced \textit{uncontrolled} relative horizontal displacements of the cavity with respect to the platform, which corresponds to an \textit{amplification} of the vibrational effect at the relatively lightweight cavity in comparison to the vibration level at the relatively heavy platform.
By implementing improvement (5), we not only increased the effective weight of the cavity by making it form a whole with its horizontal fastening structure, thus reducing the corresponding vibrational sensitivities (see the Supplementary materials), but also directly restricted uncontrolled relative horizontal displacements of the cavity to near zero with respect to the platform. Our approach shares certain features with the rigid-mounting method for realizing transportable ultra-stable cavities operating in non-laboratory environments~\cite{Leibrandt11oe, Webster11ol, Hafner20oe} and is analogous to the M\"ossbauer effect in nuclear physics~\cite{Mossbauer58Naturwissenschaften, Mossbauer64rmp}. We emphasize that the key function of improvement (5) is not to improve the ``environmental'' vibrational stability at the platform, but to suppress the vibrational noise at the cavity down to the ``environmental level'' at the platform. While the platform may still bear more vibrational noise than a base in a room-temperature, cryostat-free cavity system, the relatively heavy weight of the platform (compared with a lightweight free cavity) and its firm connection to the whole sample chamber already make the current ``environmental'' vibration instability sufficiently small to allow for $10^{-17}$-level frequency stabilities of the Si1 cavity at second-scale averaging times. Moreover, numerical simulations showed that, in the presence of a horizontal fastening structure, the effective CTE of the Si1 cavity remains dominated by single-crystalline silicon (see the Supplementary materials).  \\

\noindent\textit{3.4. Optimized Si1 frequency instability reaching $4\times 10^{-17}$ under continuous cooling}\\

To optimize the frequency stability of the Si1 cavity, we protected the whole Si1 optical path against acoustic noise, air flow, temperature drift, and other slow fluctuations by implementing improvement (6). We then performed TCH measurements and analyzed the modified Allan deviations using 27 data sets that were 100~s long for $\tau < 3$~s and 28 data sets that were 500~s long for $\tau \ge 3$~s. As shown in Fig.~\ref{fig:contiunous}a, the silicon cavity (denoted as Si1c, with improvements (1) to (6) all implemented) demonstrated $10^{-17}$-level frequency instability at $\tau = 0.9$  to 100~s, showing advantages of cryogenic silicon cavities over room-temperature cavities ULE2 and ULE3. In particular, the Si1c instability reached $6.0(3)\times 10^{-17}$ at $\tau = 1$~s and improved to $3.4(3) \times 10^{-17}$ at around 6~s. In the range of $\tau = 4$ to 12~s, we extracted an average frequency instability of $4.3(2)\times 10^{-17}$, which is 
close to the Si1 Brownian thermal noise floor at 4.9~K. At present, the one-second Si1c instability is primarily limited by residual amplitude modulation (RAM) noise (gray dashed line in Fig.~\ref{fig:contiunous}a). At short and long averaging times, the Si1c instability is  dominated by seismic vibration ($\tau < 0.1$~s) and cavity temperature fluctuation ($\tau > 40$~s) respectively; see the Supplementary materials. Furthermore, by reducing the platform temperature from 4.9~K to 3.3~K, we observed improved long-term instabilities of $4.9(8)\times 10^{-17}$ at $\tau = 100$~s and $5.5(1.2)\times 10^{-16}$ at $\tau = 1000$~s, showing the benefit of reducing the silicon CTE; see Fig.~\ref{fig:contiunous}a inset and the Supplementary materials. In accordance with the  modified Allan deviation for 4.9~K, the Si1c PSD (averaged over 140 data sets that were 100~s long) approached the Si1 Brownian thermal noise floor at Fourier frequencies of 0.04 to 0.06~Hz, showing a marked improvement over Si1b; see Fig.~\ref{fig:contiunous}b.

We investigated the Si1c ultra-narrow linewidth using two methods. In the first method, based on a systematic scheme demonstrated in Ref.~[\citenum{Bishof13prl}] and the  Si1c PSDs (140 data sets that were 100~s long), we numerically computed the minimum observable Laser1 linewidth to be $\Delta\nu_{\mathrm{Si1,I}} =9.6(3)$~mHz (median value) for a measurement time up to 100~s; see Fig.~\ref{fig:contiunous}c and the Supplementary materials. In the second method, we measured the heterodyne beat between Si1 and ULE3 (mixed down to near d.c.) with a digital oscilloscope, as shown by a sample time trace in Fig.~\ref{fig:contiunous}d. Twelve such time traces of the heterodyne beat were each analyzed via fast Fourier transform and then averaged to yield an average line profile with a fitted full width at half-maximum of $\Delta_{1,3} = 21(1)$~mHz; see Fig.~\ref{fig:contiunous}e and the Supplementary materials. Considering the typical frequency stabilities (averaged over $ 1\le \tau \le 20$~s) of Si1c and ULE3, $\bar{\sigma}_{\mathrm{Si1c}} \approx 4.5 \times 10^{-17}$ and $\bar{\sigma}_{\mathrm{ULE3}} \approx 1.6 \times 10^{-16}$ respectively, we followed an empirical approach where the linewidths of two stable lasers were added in quadrature in a beat measurement~\cite{Zhang17prl}, and extracted the Si1 contribution to the beat linewidth as $\Delta\nu_{\mathrm{Si1,II}} = \Delta_{1,3}\bar{\sigma}_{\mathrm{Si1c}}/\sqrt{\bar{\sigma}_{\mathrm{Si1c}}^2 + \bar{\sigma}_{\mathrm{ULE3}}^2} \approx 5.7(3)$~mHz, which is qualitatively consistent with $\Delta\nu_{\mathrm{Si1, I}}$. The ultra-narrow Si1 linewidth corresponds to an optical quality factor of $\nu_{_\mathrm{Si1}}/\Delta\nu_{\mathrm{Si1, I}} \approx 2.2 \times 10^{16}$ for a highly coherent optical local oscillator at 1397~nm, with $\nu_{_\mathrm{Si1}} \approx 214.6$~THz.\\

\noindent\textbf{4. Conclusion and discussion}\\

In summary, we have realized an ultra-stable 10-cm-long silicon cavity at sub-5-K temperatures and demonstrated $4.3(2)\times 10^{-17}$ laser frequency instability that approaches the cavity's Brownian thermal noise floor at 4.9~K.
Our work paves the way for enhancing the frequency stabilities of silicon cavities at several-kelvin temperatures using long cavities. With state-of-the-art ultra-low-vibration cryostat techniques~\cite{Ma23rsi} and advanced electro-optical control~\cite{Kedar24Optica}, the realization of frequency stability at the $1\times 10^{-17}$ or $10^{-18}$ level is within reach based on a silicon cavity with a length of 20~cm or more. Such highly stable optical local oscillators hold promise for enabling ground- and space-based optical atomic clocks with $10^{-19}$ to $10^{-21}$-level stabilities and other precision interferometric systems, which in turn will provide ultra-sensitive quantum sensors for fundamental science including  gravitational waves and dark matter~\cite{Kolkowitz16prd, Tino19epjd, Cole23apl}.
 The CQQM method also provides an efficient way to investigate fundamental and electro-optical frequency noise for various new mirror coatings.\\

\noindent\textbf{Conflict of interest}\\

The authors declare that they have no conflict of interest.\\

\noindent\textbf{Acknowledgments}\\


We are grateful to Yige Lin, Yu-Ao Chen, Han-Ning Dai, Ping Xu, Xi Lin for very insightful discussions. We thank Jian-Wei Pan, Jian-Yu Wang, Long-Sheng Ma, Shou-Gang Zhang, Ke-Lin Gao, Yan-Yi Jiang, Hai-Feng Jiang, Jie Zhang, Zhan-Jun Fang, Su Fang,  Xiaoji Zhou, Bo Song, Xing-Yang Cui, Shao-Shuai Liu, Yan-Yang Jiang,  Ye Li, Xian-Qing Zhu, Wen-Shuai Peng, Tangyin Liao, Bingkun Lu, Qunyi Ouyang, Rui Wu, Shize Du, Tao Deng, Yi-Lin Wu, Shiyin Kuang for discussions and technical support. This work was supported by the Chinese Academy of Sciences Strategic Priority Research Program (XDB35020100), the Hefei National Laboratory, and the Innovation
Program for Quantum Science and Technology
(2021ZD0301903).  \\

\noindent\textbf{Author contributions}\\

Xibo Zhang conceived the project.  Zhi-Ang Chen, Hao-Ran Zeng, Wen-Wei Wang, Run-Qi Lei performed the experiments. Zhi-Ang Chen, Wen-Wei Wang, Han Zhang performed numerical simulations for designing the silicon cavity and associated apparatus. Jian-Zhang Li and Cai-Yin Pang performed the manufacturing and cleansing of the spacer, as well as the optical contacting of the silicon mirrors to the spacer. She-Song Huang contributed important insights for suppressing optical contamination based on high-vacuum environment and for upgrading the cavity support structure to improve the overall rigidity.
All authors contributed to the data analysis and the writing of this manuscript.\\

\newcommand{\pzcS}{\mathpzc{S}}
\newcommand{\pzci}{\mathpzc{i}}

\renewcommand{\theequation}{S\arabic{equation}}
\renewcommand{\thefigure}{S\arabic{figure}}
\renewcommand{\thetable}{S\arabic{table}}

\newcommand{\f}{|f\rangle}
\newcommand{\e}{|e\rangle}
\newcommand{\fth}{|f;\,+\frac{3}{2}\rangle}
\renewcommand{\eth}{|e;\,+\frac{3}{2}\rangle}
\renewcommand{\theequation}{S\arabic{equation}}
\renewcommand{\thefigure}{S\arabic{figure}}

\newcommand{\boldrho}{\mbox{\boldmath$\rho$}}
\newcommand{\bg}[1]{\mbox{\boldmath$#1$}}
\definecolor{red}{rgb}{0.7,0,0}
\definecolor{green}{rgb}{0.,0.35,0.}
\definecolor{blue}{rgb}{0.2,0.2,0.7}
\definecolor{black}{rgb}{0.15,0.15,.15}
\newcommand{\com}[1]{{\color{blue}\small\   #1 }}
\newcommand{\ad}{a^{\dagger}}
\newcommand{\al}{\alpha^{\dagger}}
\newcommand{\be}{\beta^{\dagger}}
\newcommand{\ga}{\gamma^{\dagger}}
\newcommand{\de}{\delta^{\dagger}}
\newcommand{\fd}{f^{\dagger}}
\newcommand{\an}{\mathcal N}
\newcommand{\bfn}{\mathbf n}
\newcommand{\one}{\mbox{$1 \hspace{-1.0mm}  {\bf l}$}}
\newcommand{\spec}{\mbox{spec}}
\newcommand{\bracket}{\rangle \langle}
\newcommand{\vac}{|0\rangle }
\newcommand{\ket}[1]{\left|#1\right\rangle}
\newcommand{\bra}[1]{\left\langle#1\right|}
\newcommand{\rem}[1]{\textbf{\textcolor{red}{[#1]}}}

\newcommand{\ree}{\rho_{ee}^{mm}}
\newcommand{\rgg}{\rho_{gg}^{mm}}
\newcommand{\reg}{\rho_{eg}^{mm}}
\newcommand{\rge}{\rho_{ge}^{mm}}
\clearpage

\section*{SUPPLEMENTAL MATERIAL}

	\section{Experimental setup and methods}

	We design and realize a high-finesse silicon optical cavity that is relatively long compared to those in Refs.~\cite{Zhang17prl,Robinson19optica,Kedar23optica}. The silicon cavity spacer and mirror substrates are manufactured by Beijing Xinnan Zhike Optoelectronic Technology and are made based on a high-purity monocrystalline silicon ingot grown by the float-zone method, with a silicon purity better than $99.999999999\%$. The spacer is about 10~cm long, with an optical axis oriented along the crystalline $\langle 111 \rangle$ axis of silicon. {\color{black}A set of three venting holes are designed to be parallel to the crystalline $\langle 10\bar{1} \rangle$, $\langle \bar{1}10 \rangle$ and $\langle 0\bar{1}1 \rangle$ axes of silicon respectively, and to be centered in the same plane perpendicular to the $\langle 111 \rangle$ axis. Thus, these three venting holes are symmetrically distributed on the body of the spacer, with relative azimuthal positions separated by $120^{\circ}$.} A pair of high-reflectivity cavity mirrors, with $\infty$ and 1-m radii of curvature respectively, are superpolished by Coastline Optics, then coated with dielectric coating technique by FiveNine Optics, and subsequently optically contacted to the spacer by Beijing Xinnan Zhike Optoelectronic Technology. The high-reflectivity mirror surfaces are perpendicular to the $\langle 111 \rangle$ axis. Furthermore, the $\langle 10\overline{1} \rangle$ crystalline axes of the mirrors and the spacer are aligned to an accuracy of one degree.

	For the laser stability measurements in this work, we couple Laser1 well into the Si1 TEM${}_{00}$ mode.  By assuming the two cavity mirrors have the same coefficients for reflection ($R_{\mathrm{M}}$), transmission ($T_{\mathrm{M}}$), and loss ($L_{\mathrm{M}}$), we can estimate these coefficients based on the measurements of the cavity finesse, optical power transmission through the cavity, and the interference contrast in the Pound-Drever-Hall (PDH) locking; see Section~\ref{subsec:estimateRTL} for details. During the cooling, we monitor the carrier transmission and PDH interference contrast, and observe that they remain stable within measurement uncertainties.

	The 10-cm-long single-crystalline silicon cavity is installed in a deeply customized ``cryostat sample chamber'' that is connected to a ``cryostat cooler chamber'', with both chambers manufactured by Montana Instruments. The cryostat can reach as low operating temperature as about 3~K; see Ref.~\cite{Wang23fip} for details. During the cooling process, the variation rate of Laser1 frequency has changed its sign twice during the cooling, which is consistent with the two zero-crossing points (near 124~K and 16~K respectively) of the silicon coefficient of thermal expansion (CTE). 
	After the cryostat platform temperature reached about 3~K,
	we observed a drift rate below 100~mHz/s for the Laser3-Laser1 beat frequency,
	which is again consistent with the asymptotically vanishing silicon CTE as the temperature approaches zero. 
	Under the lowest platform temperature of about 3~K, we measure a Si1 free spectral range of 1.532~GHz and hereby calibrate an accurate cavity length of 9.79~cm.

	At the same time, the apparatus in Ref.~\cite{Wang23fip} is insufficient for realizing a few-kelvin silicon cavity with $10^{-17}$-level frequency stability. For example, an earlier version of the silicon cavity (Si0) shows subtle evidences of contamination: with 2-Torr starting pressure before the cooling, Si0 exhibits substantially-lower-than-Si1 and spatially inhomogeneous finesse at few-kelvin platform temperatures ($2.0\times 10^5$ for the TEM${}_{00}$ mode and $2.8\times 10^5$ for TEM${}_{01}$), which also degrades (to $2.3\times 10^5$ for TEM${}_{01}$) after nine-month cryogenic operation.

	As mentioned in the main text, in this work we further implement six technical improvements with respect to the cryostat in Ref.~\cite{Wang23fip} and the Si0 cavity. Below we provide supplementary information on these improvements.

	\subsection{Preparation of a clean high-vacuum environment for the silicon cavity} 
	This section refers to improvements (1) and (2). 
	
	\textit{On improvement (1).---}The purpose of improvement (1) is to eliminate obviously harmful outgassing processes. To fulfill this purpose, 
	we must choose proper high-vacuum-compatible thermally conducting material that is applied between surfaces of adjacent components. Specifically, we abandon the Apiezon N grease (previously applied in Ref.~\cite{Wang23fip}), which is non-high-vacuum-compatible but remains the standard practice by the cryostat manufacturer Montana Instruments. In fact, after an earlier complete cooling-down and warming-up thermal cycle, we find small patches of yellowish mark at some in-vacuum surfaces, which is likely caused by the use of Apiezon N grease and the associated outgassing. 
	
	To eliminate such harmful outgassing, we apply the following high-vacuum-compatible method: (1) thoroughly cleaning all in-vacuum surfaces (including the cavity spacer) sequentially using laundry detergent solution, d-limonene, acetone, and ethanol, and then (2) attaching pieces of indium metal film with 10-$\mu$m thickness between surfaces of adjacent
	components to improve thermal contact.\\
	
	\textit{On improvement (2).---}The purpose of improvement (2) is to significantly suppress the contamination of cavity mirrors due to condensation of gas molecules on the mirrors. We abandon Montana Instruments' standard practice of starting the cooling as soon as the vacuum pressure drops below 2 Torr because under this level of starting pressure, a lot of gas molecules can still condense onto the cavity mirrors during the cooling process and thus contaminate the high-reflectivity coating. Instead, we permanently install an ion pump (Gamma Vacuum, 45~L/s) and a full-range vacuum gauge (Agilent Technologies, FRG-700) to the sample vacuum chamber. Based on two cascaded molecular-turbo pumps, we reduce the pressure to about $2\times 10^{-6}$~Torr. We then start the ion pump, seal an all-metal angle valve (such that the system relies on the ion pump alone), and shut down the molecular-turbo pumps. The final pressure (read at the ion pump and confirmed by the vacuum gauge) reaches about $2\sim 4\times 10^{-7}$~Torr under room temperature before any cooling starts, which is likely limited by the outgassing from more than ten Viton O-rings in the cryostat vacuum chamber.

	\subsection{Mechanical structure for vertically supporting and horizontally fastening the silicon cavity}
	This section refers to improvements (3) and (5). 
	
	\textit{On improvement (3).---}The purpose of improvement (3) is to improve rigidity of the mechanical structure that supports the silicon cavity in the vertical direction.  In the building of our initial three-point supporting structure made of G10 and stainless steel (similar to that in Ref.~\cite{Zhang17prl}), Montana Instruments made a \textbf{very subtle but severe} mistake by wrongly implementing three separate small rods that were inserted into three holes on the top surface of the supporting structure, which didn't realize close fitting and had rather weak rigidity within the whole supporting structure, thus leading to uncontrolled frequency sensitivity to outer vibrational perturbations. 
	
	To correct this mistake, we build a new, semi-monolithic, all-stainless-steel supporting structure that has no shakable components, which thus eliminates the aforementioned uncontrolled vibrational sensitivity and corresponds to higher mechanical resonance frequencies (about 600~Hz and higher, based on finite-element analysis). Here, stainless steel is chosen because of its combined properties of rigidity and poor thermal conductivity under few-kelvin temperatures. Our new implementation is based on an independent design for improving the overall rigidity within the supporting structure while still maintaining its low thermal conductivity. Later we become aware that our implementation shares certain features with that in Ref.~\cite{Kedar23thesis}. 
	
	\textit{On improvement (5).---} The purpose of improvement (5) is to reduce the frequency instability induced by cryostat cooler vibration, in particular frequency instability induced by horizontal vibrational noise. \textbf{The key idea is to implement three-point fastening of the Si1 cavity horizontal position, and at the same time use as little material as possible in realizing the horizontal fastening such that not too much force is exerted onto the cavity.} In the present implementation, we choose to realize such horizontal fastening via three small horizontal cylinders with 5-mm diameters, whose main axes are $120^\circ$ with respect to each other, as marked by the three blue arrows in Fig.~\ref{fig:sm_horFastenStrRelief} as well as Fig.~1a of the main text. The height of centers of these three small cylinders is equal to the center height of the silicon cavity. These three horizontal cylinders are parts of a monolithic Polyether-ether-ketone (PEEK) structure (beige color in Fig.~\ref{fig:sm_horFastenStrRelief}). Here, PEEK material is chosen for its relatively small Young's modulus compared to single-crystalline silicon. The PEEK structure itself is fixed onto a stainless steel supporting structure under it, while the stainless steel structure is in turn fixed onto a base made of oxygen-free high-conductivity copper (shown by goldish yellow color in Fig.~\ref{fig:sm_horFastenStrRelief}).
	
	\textit{Symmetry considerations for improvement (5).---}We note that in the implementation of improvement (5), two careful considerations are taken into account regarding the symmetry of the three-point fastening structure. In the vertical direction, the center heights of the three small cylinders are chosen to be the same as the center height of the silicon cavity. In the horizontal directions, the azimuthal positions of the symmetrically distributed three fastening points are designed to align with those of the three symmetrically positioned venting holes on the cavity. In this configuration, we observe in finite-element analysis that the optimum relative angle of the three vertical supporting points (with respect to three venting holes) remains roughly the same (near zero degree) without and with improvement (5). Thus, in practice, we choose this relative angle to be zero degree to facilitate the installation.
	
	Under a platform temperature around 4.9~K, we measure the vibrational sensitivities of the Si1 cavity before and after implementing improvement (5). The results are listed in Table~\ref{table:vibsen}. We observe that, from Si1a to Si1b, all three vibrational sensitivities reduce substantially, with $k_{\mathrm{H}_{1,2}}$ both reaching the $10^{-11}/g$ regime, which is consistent with a picture of increased effective mass for the cavity in the presence of horizontal fastening, as explained in Sec.~3.3 of the main text.

	\textit{Close fitting and strain relief.---}In order to realize an almost-close-fit configuration between the cavity and the horizontal fastening structure, we have made more than 20 copies of the PEEK structure with slightly different parameters, picked out one structure that matches the cavity outer diameter in a closest manner. Finally, in order to further improve the mechanical matching and to protect the silicon cavity from excessive stress applied by hard objects, we apply strain relief by adding a single layer of relatively soft \textit{indium metal film} with thickness of 10~$\mu$m 
	onto the surfaces of these three small cylinders used to fasten the position of the silicon cavity in the horizontal directions; as illustrated in Fig.~\ref{fig:sm_horFastenStrRelief}. We also add such indium metal film on the surfaces of the three hemispheres supporting the silicon cavity in the vertical direction.
	
	\setcounter{table}{0}
	\setcounter{figure}{0}
	
	\begin{table}
		\center
		\caption{Vibrational sensitivities of the Si1 cavity along three spatial directions, measured around 4.9~K platform temperature, before and after the implementation of improvement (5). Here, $k$ denotes the vibrational sensitivity, $g$ is the gravitational acceleration, and the three spatial directions are horizontal 1 ($\mathrm{H}_1$) , horizontal 2 ($\mathrm{H}_2$) and vertical (V).}
		\label{table:vibsen}
		\begin{tabular}{lcccc} 
			\specialrule{1.0pt}{0pt}{0pt}
			Si1 cavity  & $k_{\mathrm{H}_1}$   & $k_{\mathrm{H}_2}$ & $k_{\mathrm{V}}$\\
			configuration & & Unit of $k$: $10^{-11}/g$ & \\
			\specialrule{0.5pt}{0pt}{0pt}
			Si1a && with improvements (1) to (4) &\\
			& $6.6$  & $10$  & $27$ \\
			\hline
			Si1b & & with improvements (1) to (5)  & \\
			&  $3.8$ & $4.8$ & $17$ \\
			\hline
			Si1c & & with improvements (1) to (6)& \\
			&  & Same as Si1b  &  \\
			\specialrule{1.0pt}{0pt}{0pt} 
		\end{tabular}
	\end{table}

	\begin{figure}
		\centering
		\includegraphics[scale=0.65]{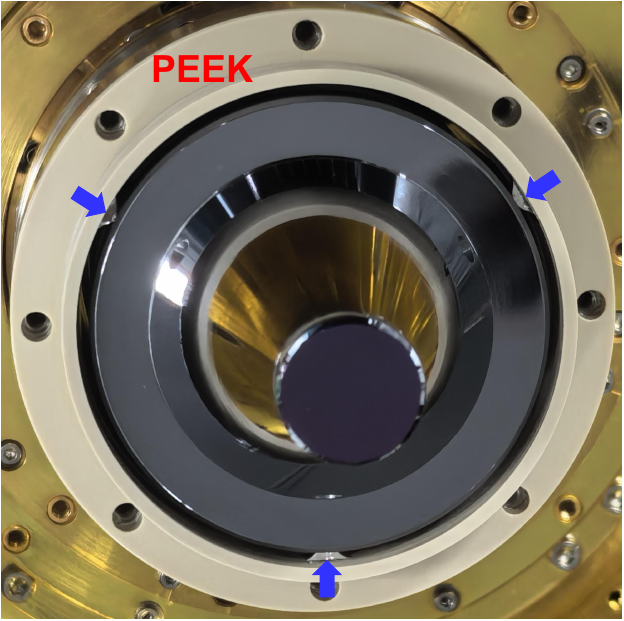}
		\caption{Illustration of the horizontal fastening for Si1 cavity and strain relief based on indium metal film. This supplemental figure provides an enlarged and marked view of the upper inset of Fig.~1a of the main text. A monolithic PEEK structure (beige structure marked by red letters) surrounds the Si1 cavity and ``touches'' the cavity via three small horizontal cylinders (marked by the three blue arrows), which thus fasten the position of the silicon cavity in the horizontal direction. The PEEK structure itself is fixed onto a stainless steel supporting structure under it, while the stainless steel structure is in turn fixed onto a base made of oxygen-free high-conductivity copper (with goldish yellow color). To provide strain relief and further protect the cavity, we add a single layer of relatively soft indium metal film with 10-$\mu$m thickness onto the surface of each small cylinder. 				
		}
		\label{fig:sm_horFastenStrRelief}
	\end{figure}

	\subsection{Three-cornered-hat measurement}
	In our experiment, there are three ultra-stable lasers, namely the silicon cavity laser (``LoongLaser-I'', or simply ``Laser1'') and two reference lasers (Laser2 and Laser3), which are phase-locked to three ultra-stable cavities Si1, ULE2, and ULE3 respectively. The three-laser composite system enables us to measure optical beat frequencies (using K+K FXE frequency counters) and perform three-cornered-hat frequency stability analysis to determine the modified Allan deviation and noise power spectral density for each laser~\cite{Premoli93IEEE,Zhang17prl}. As an example, for  a $1/f$ noise with power spectral density of $1.71\times 10^{-33}/f$, with $f$ being the Fourier frequency in Hz, the  modified Allan deviation is $4.0\times 10^{-17}$, which corresponds to a standard Allan deviation of $4.9\times 10^{-17}\approx 5\times 10^{-17}$, as also noted in Refs.~\cite{Dawkins07IEEE, Matei17prl}.
	
	Laser1 is realized by phase-locking a commercial laser (Toptica DL Pro, operated near 1396.8892~nm) to the TEM${}_{00}$ mode of Si1 with a 600-kHz locking bandwidth and controlled relatively low incident optical power (1.1~$\mu$W before the entrance viewport of the vacuum chamber). This incident power is controlled at a sufficiently low level such that (1) the laser leads to negligible heating of the related cryogenic components and (2) a realistic fractional power instability at the $10^{-4}$ level can correspond to a frequency instability that is substantially smaller than the Brownian thermal noise floor (see the hexagons in Fig.~\ref{fig:sm_indivinoise}). At the same time, this incident power is still large enough such that the photon shot noise and the technical noise of the PDH photo-detector give rise to frequency instabilities (here at the levels of $7\times 10^{-18}$ at 1~s and $9\times 10^{-18}$ at 1~s respectively) that are well below the Brownian thermal noise floor. For this PDH locking, a free-space resonant electro-optic phase modulator (Qubig, PM7-SWIR) is chosen for performing phase modulation and a high-sensitivity commercial free-space InGaAs avalanche photodetector (Thorlabs APD430C/M) is used for measuring the raw error signal.
	
	As illustrated by Fig.~1b of the main text, the frequencies of the 698-nm Laser2 and 1397-nm Laser3 are compared by first doubling the Laser3 frequency by a single-pass periodically poled lithium niobate (PPLN) waveguide (Shandong Jiliang Information Technology, SHG) and then recording the beat of two 698-nm lasers. The additional frequency noise introduced in the frequency doubling process by the  single-pass PPLN waveguide has been experimentally tested to be at the level of $1\times 10^{-17}$ or smaller for averaging times above 0.5 second. In this work, the optical beats are measured using dead-time-free K+K FXE frequency counters in phase-averaging mode, which corresponds to a lambda-mode counting scheme~\cite{Hagemann13thesis}.
	
	Using synchronized K+K FXE frequency counters, we measure two raw signals of optical beat frequencies:
	\begin{itemize}
		\item $\nu_{_\mathrm{3-Si}} \equiv \nu_{_\mathrm{ULE3}} - \nu_{_\mathrm{Si1}}$ between the ULE3 cavity and the Si1 cavity,
		\item $\nu_{_\mathrm{3-2}} \equiv 2\nu_{_\mathrm{ULE3}} - \nu_{_\mathrm{ULE2}}$ between the ULE3 cavity (after frequency doubling) and the ULE2 cavity,
	\end{itemize}   where $\nu_{_\mathrm{Si1}}$, $\nu_{_\mathrm{ULE2}}$ and $\nu_{_\mathrm{ULE3}}$ are the frequencies of Laser1, Laser2, and Laser3, and the ``2'' on the right-hand-side of the equation defining $\nu_{_\mathrm{3-2}}$ represents the frequency doubling via PPLN waveguide. With these two measured beat frequencies, the third beat frequency between the ULE2 and Si1 cavities are computed via a linear combination of $\nu_{_\mathrm{3-Si}}$ and $\nu_{_\mathrm{3-2}}$:
	\begin{eqnarray}
		\nu_{_\mathrm{2-Si}} & \equiv & \nu_{_\mathrm{ULE2}} - 2\nu_{_\mathrm{Si1}}  \nonumber\\
		& = & 2\nu_{_\mathrm{3-Si}} - \nu_{_\mathrm{3-2}},
	\end{eqnarray}
	as similarly performed in Ref.~\cite{Matei17prl} because the beat measurement system does not introduce appreciable additional noise. In fact, for investigating the counters' background noise level, we feed the RF signals from commercial frequency synthesizers into the K+K FXE counters, and observe that the measured noise floor corresponds to a ``fractional frequency instability'' at the $1\times 10^{-19}$ level (derived using the same optical carrier frequency at 1397~nm) at one-second averaging time, which further reduces at longer averaging times.
	
	Assuming uncorrelated noise between three ultra-stable lasers, the frequency instability of an individual laser at a given time can be extracted as 
	\begin{eqnarray}\label{eq:tch_mADEV}
		\textrm{mod }\sigma_{_\mathrm{1}} & = & \frac{1}{2}\sqrt{\frac{4 \textrm{mod }\sigma^2_{\nu_{_\mathrm{3-Si}}} + \textrm{mod }\sigma^2_{\nu_{_\mathrm{2-Si}}}- \textrm{mod }\sigma^2_{\nu_{_\mathrm{3-2}}}}{2}} \nonumber\\
		\textrm{mod }\sigma_{_\mathrm{2}} & = & \sqrt{\frac{\textrm{mod }\sigma^2_{\nu_{\mathrm{3-2}}} + \textrm{mod }\sigma^2_{\nu_{\mathrm{2-Si}}}- 4\textrm{mod }\sigma^2_{\nu_{\mathrm{3-Si}}}}{2}} \nonumber\\
		\textrm{mod }\sigma_{_\mathrm{3}} & = & \frac{1}{2}\sqrt{\frac{4 \textrm{mod }\sigma^2_{\nu_{_\mathrm{3-Si}}} + \textrm{mod }\sigma^2_{\nu_{_\mathrm{3-2}}}- \textrm{mod }\sigma^2_{\nu_{_\mathrm{2-Si}}}}{2}}, \nonumber\\
	\end{eqnarray}
	where $\textrm{mod }\sigma_{_\mathrm{1}}\equiv \textrm{mod }\sigma_{_\mathrm{Si1}}$, $\textrm{mod }\sigma_{_\mathrm{2}} \equiv \textrm{mod }\sigma_{_\mathrm{ULE2}}$, and $\textrm{mod }\sigma_{_\mathrm{3}} \equiv \textrm{mod }\sigma_{_\mathrm{ULE3}}$ are the modified Allan deviations of $\nu_{_\mathrm{Si1}}$, $\nu_{_\mathrm{ULE2}}$, and $\nu_{_\mathrm{ULE3}}$ respectively. Accordingly, $\textrm{mod }\sigma_{\nu_{_\mathrm{3-Si}}}$, $\textrm{mod }\sigma_{\nu_{_\mathrm{3-2}}}$, and $\textrm{mod }\sigma_{\nu_{_\mathrm{2-Si}}}$ are the modified Allan deviations of the corresponding three beat frequencies $\nu_{_\mathrm{3-Si}}$, $\nu_{_\mathrm{3-2}}$, and $\nu_{_\mathrm{2-Si}}$ respectively. Each frequency instability $\textrm{mod }\sigma_{_\mathrm{1,2,3}}$ is normalized by the corresponding optical frequency of Laser1/Laser2/Laser3 to yield a fractional frequency instability, which is also referred to as ``fractional frequency stability'' in the main text. Here, each modified Allan deviation is a function of averaging time $\tau$. In case there is need of removing possible correlations among these lasers, we apply an extended technique described in Ref.~\cite{Premoli93IEEE} for computing the modified Allan deviations.
	
	\textbf{Data averaging and statistical methods for Figure~4 of the main text.---}For Fig.~4a of the main text, we use 27 data sets that are 100~s long for $\tau < 3$~s and 28 data sets that are 500~s long for $\tau \ge 3$~s. For Fig.~4a inset, we use 25 data sets that are 100~s long for $\tau < 3$~s, 21 data sets that are 500~s long for $3 \le \tau \le 100$~s, 17 data sets that are 1000~s long for $100 < \tau \le 200$~s, and 6 data sets that are 5000~s long for $200 < \tau \le 1000$~s. For Fig.~4b and 4c of the main text, we use 140 data sets that are 100~s long. For Fig.~4e, we use 12 data sets that are 100~s long and similar to the time trace in Fig.~4d.
	
	Each single data set for optical beat frequency measurements (using a K+K FXE counter) has an independent linear drift removed. Then, taking Fig.~4a as an example, we determine the modified Allan deviation and the corresponding $1\sigma$ statistical uncertainty of each single data set based on the statistical method of Ref.~\cite{Riley08NIST}. Subsequently, the individual results based on multiple data sets under the same experimental condition are averaged using a conservative statistical method of weighted averaging, with the $1\sigma$ statistical uncertainty inflated by the square root of the reduced chi-squared~\cite{Bloom14nature}.

	\textbf{Estimated frequency drift rate for the Si1 cavity.---}Here we estimate the low drift rate of the Si1 cavity based on existing literature on sub-5-K silicon cavities~\cite{Robinson19optica, Robinson23thesis} and proper scaling with respect to the experimental parameter. In Ref.~\cite{Robinson19optica}, it has been demonstrated that a 6-cm-long cryogenic silicon cavity operating at 4~K has a fractional frequency drift rate that is roughly proportional to the transmitted optical power through the cavity, with a proportionality coefficient of about $7\times 10^{-21}$/s/nW. In Ref.~\cite{Robinson23thesis}, it has been further shown that the frequency drift rates are similar under operating temperatures of 3.8~K, 4.9~K, and 16~K for the silicon cavity. Compared to Ref.~\cite{Robinson19optica}, our Si1 cavity has a transmitted optical power of about 230~nW, and its operating temperature of 4.9~K is also similar to the value (4~K) in Ref.~\cite{Robinson19optica}. Considering (1) the high purity ($99.999999999\%$) of the silicon material we use, (2) the high superpolishing quality by Coastline Optics, (3) the high quality of dielectric coating by FiveNine Optics, and (4) that both the Si1 cavity spacer and mirror substrates were cut from the same ingot and had aligned crystalline axes, we judge that our Si1 cavity should have  high optical quality similar to that of the silicon cavity in Refs.~\cite{Robinson19optica, Robinson23thesis}. Therefore, we estimate the frequency drift rate of the Si1 cavity based on the proportionality coefficient in Ref.~\cite{Robinson19optica} and our transmitted power of 230~nW, yielding a fractional frequency drift rate of  $7\times 10^{-21}$/s/nW$\times 230$~nW$=1.6\times 10^{-18}$/s. At 1397~nm, this fractional drift rate corresponds to an absolute frequency drift rate of 0.00034~Hz/s $<0.4$~mHz/s, which is well below 1~mHz/s.

\begin{figure}
		\centering
		\includegraphics[scale=0.45]{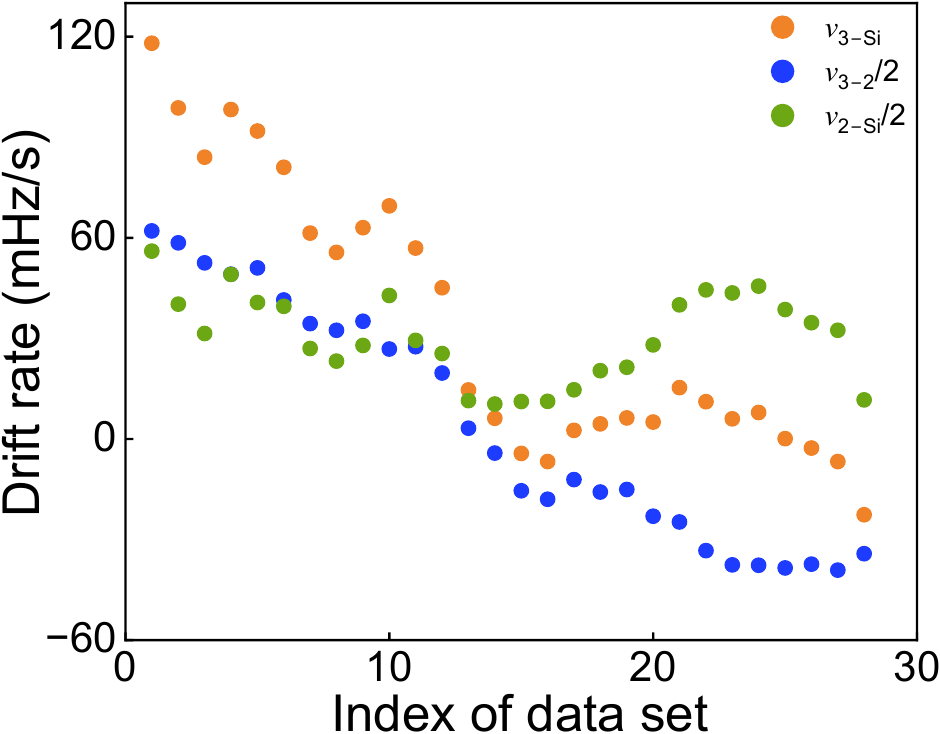}
		\caption{Drift rates of optical beat frequencies. Here, the optical beat frequencies are measured in 28 data sets for Fig.~4a of the main text, where each data set is 500 seconds long. In addition, the optical beat frequencies are converted into  equivalent values for comparing frequencies of 1397-nm lasers, which corresponds to $\nu_{_\mathrm{3-Si}}$, $\frac{1}{2}\nu_{_\mathrm{3-2}}$, and $\frac{1}{2}\nu_{_\mathrm{2-Si}}$.
		}
		\label{fig:sm_DriftRates}
	\end{figure}

	\textbf{Measured frequency drift rates of the optical beats.---} We investigate the typical drift rate of the optical beat frequencies measured in 28 data sets that are 500 seconds long for Fig.~4a of the main text. Here, we measure the drift rates of the three beats of $\nu_{_\mathrm{3-Si}}$, $\frac{1}{2}\nu_{_\mathrm{3-2}}$, and $\frac{1}{2}\nu_{_\mathrm{2-Si}}$ and plot these drift rates in Fig.~\ref{fig:sm_DriftRates}, where we have converted all beat frequencies to the equivalent values for comparing frequencies of 1397-nm lasers. For a certain beat frequency, we fit the drift rate for each data set, and take the individual absolute value, and then perform averaging over the 28 data sets. The resultant mean drift rates for $\nu_{_\mathrm{3-Si}}$, $\frac{1}{2}\nu_{_\mathrm{3-2}}$, and $\frac{1}{2}\nu_{_\mathrm{2-Si}}$ are 37~mHz/s, 31~mHz/s, and 30~mHz/s respectively, which are about two orders of magnitude larger than the estimated Si1 drift rate and therefore likely dominated by the drift rates of two room-temperature ULE cavities. We also note that, from our operating experience of an ULE-cavity-based 689-nm ultra-stable laser in ultracold strontium atoms experiments, the daily frequency drift of a ULE cavity is on the order of several kilohertz per day (based on daily correction of the laser frequency using a single-frequency, second-stage strontium magneto-optical trap). Taking a 5~kHz/day drift rate as a typical value, we compute that such a daily drift corresponds to an average drift rate of about 60~mHz/s for a laser at 689~nm, which is fairly consistent with the level of the measured mean drift rates of the optical beats (converted to values for beats between 1397-nm lasers) extracted from data sets that are 500 seconds long.

	\subsection{Estimating the mean reflection, transmission, and loss coefficients for the cavity mirrors} \label{subsec:estimateRTL}
	By assuming the two cavity mirrors have the same coefficients (defined for optical power) of reflection ($R_{\mathrm{M}}$), transmission ($T_{\mathrm{M}}$), and loss ($L_{\mathrm{M}}$), we can estimate these coefficients based on the measurements of three physical quantities: the cavity finesse $\mathcal{F}$, the on-resonance optical power transmission $\mathcal{T}$ through the cavity and the sample chamber, and the on-resonance fractional reflectance $\mathcal{R}_{\mathrm{PDH}}$ based on the voltage of the PDH-locking photo-detector. The cavity finesse is measured by ring-down measurements illustrated in the inset of Fig.~1c of the main text, and is related to $R_{\mathrm{M}}$ via the following expression~\cite{Nagourney14QEbook}:
	\begin{eqnarray}\label{eq:RTL1}
		\mathcal{F} & = & \frac{\pi \sqrt{R_{\mathrm{M}}}}{1-R_{\mathrm{M}}}.
	\end{eqnarray} Below we describe the factors that contribute to $\mathcal{T}$ and $\mathcal{R}_{\mathrm{PDH}}$.
	
	The on-resonance optical power transmission $\mathcal{T}$ is determined as the ratio of the optical power that comes out of the sample vacuum chamber and the total optical power of the laser beam that enters the sample vacuum chamber. Besides $R_{\mathrm{M}}$, $T_{\mathrm{M}}$, and $L_{\mathrm{M}}$, several other experimental factors affect the value of $\mathcal{T}$. First, on and inside the sample vacuum chamber, there are four N-BK7 glass windows on each side of the cavity (namely, eight windows in total). These eight N-BK7 windows are measured to have a total transmission of $T_{\mathrm{BK7}} \approx 81\%$ for our laser at 1397~nm. Second, we measure that for the incident laser beam, about $\beta \approx 70\%$ of the optical power is in the on-resonance  carrier frequency component, and about $30\%$ is in the off-resonance modulation sidebands. Third, the coupling efficiency $\eta$ of the incident laser beam into the TEM${}_{00}$ mode of the cavity may be less than $100\%$ in the real apparatus with experimental imperfections. Taking these factors into account, the measured raw transmission  $\mathcal{T}$ can be modeled as follows~\cite{Nagourney14QEbook,Martin13thesis}:
	\begin{eqnarray}\label{eq:RTL2}
		\mathcal{T} & = & T_{\mathrm{BK7}} \beta \eta \left(\frac{T_{\mathrm{M}}}{T_{\mathrm{M}} + L_{\mathrm{M}}}\right)^2.
	\end{eqnarray}
	
	The on-resonance fractional reflectance $\mathcal{R}_{\mathrm{PDH}}$ for the PDH locking is also affected by the values of $\beta$ and $\eta$. The measured raw signal can be modeled as follows~\cite{Martin13thesis, Nagourney14QEbook}:
	\begin{eqnarray}\label{eq:RTL3}
		\mathcal{R}_{\mathrm{PDH}} & = & \beta\eta\left(\frac{L_{\mathrm{M}}}{T_{\mathrm{M}}+L_{\mathrm{M}}}\right)^2  + \beta (1-\eta) + (1-\beta), \nonumber \\
	\end{eqnarray}
	where on the right-hand side, the first term represents the contribution of coherent interference between on-resonance reflected electric fields, the second term represents the incoherent contribution by the on-resonance frequency component that is not coupled into the cavity due to imperfect spatial mode matching between the incident laser beam and the cavity, and the third term represents the incoherent contribution by off-resonance modulation sidebands. To provide a reference value, for a fully off-resonance incident laser beam, one has $\mathcal{R}_{\mathrm{PDH},\textrm{off.res}} = 100\%$. 
	
	Under the lowest platform temperatures of about 3~K, we measure a set of typical parameters to be $\mathcal{F} \approx 3.66\times 10^5$, $\mathcal{T} \approx 20.7\%$, and $\mathcal{R}_{\mathrm{PDH}} \approx 49.5\%$. Based on Eqs.~\ref{eq:RTL1}, \ref{eq:RTL2}, and \ref{eq:RTL3}, we solved the mean reflection coefficient, mean transmission coefficient, and mean loss coefficient as 
	\begin{eqnarray}
		R_{\mathrm{M}} & \approx & 0.9999914\nonumber\\
		T_{\mathrm{M}} & \approx & 5.8~\mathrm{ppm}\nonumber\\
		L_{\mathrm{M}} & \approx & 2.8~\mathrm{ppm}\nonumber
	\end{eqnarray}
	and a mode coupling efficiency $\eta = 80.8\%$. 
	
	Finally, to be more conservative in describing the optical quality of our cavity mirrors, we round $T_{\mathrm{M}}$ down to the previous integer and round $L_{\mathrm{M}}$ up to the next integer, namely $T_{\mathrm{M}} \approx 5$~ppm and $L_{\mathrm{M}} \approx 3$~ppm, as stated in Section~3.1 of the main text.

	\section{Determination of ultra-narrow linewidth}
	
	\subsection{Linewidth computation based on measured noise power spectral density}

	\begin{figure}
		\centering
		\includegraphics[scale=0.35]{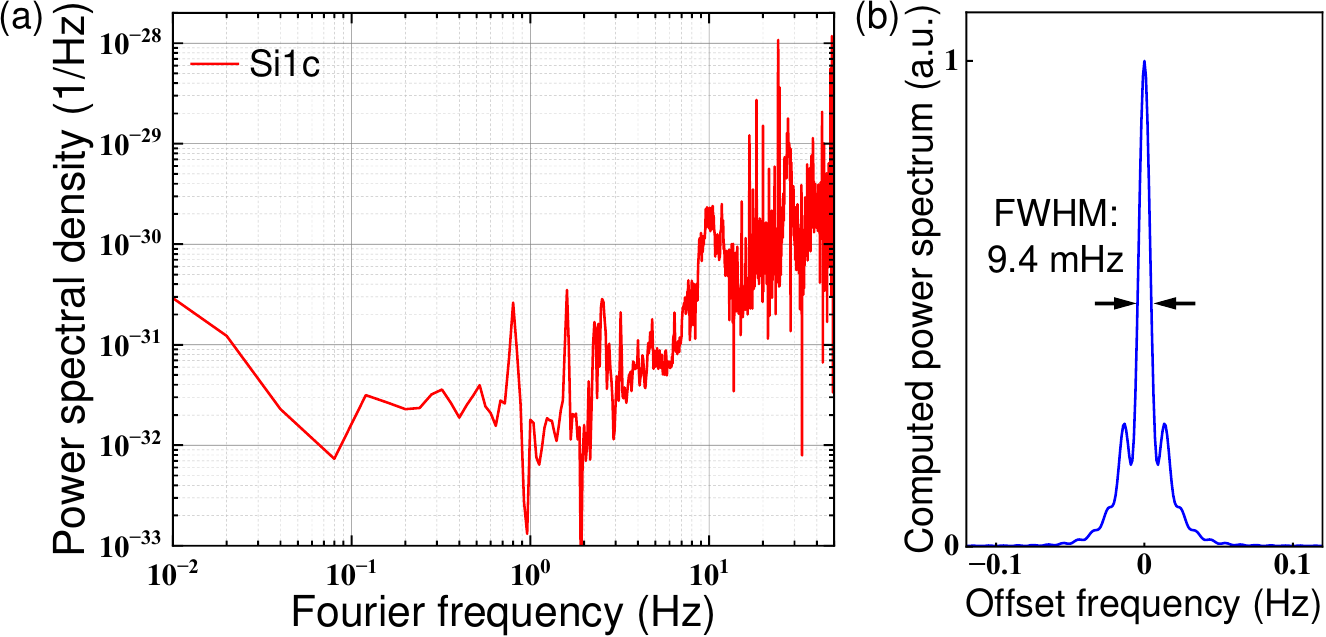}
		\caption{One individual sample data set for the single-sided frequency noise power spectral density of Laser1 (with the silicon cavity in the Si1c condition) and the accordingly computed line profile. (a) Single sided frequency noise power spectral density. (b) Computed power distribution $S_{E,\mathrm{Si1}}(\delta\nu)$ of Laser1's electric field (as a function of the offset frequency $\delta\nu$), with its full width at half-maximum (FWHM) extracted. The numerical computation is based on a systematic methodology in Ref.~\cite{Bishof13prl}.
		}
		\label{fig:sm_compFWHMlinewidth}
	\end{figure}

	If the modified Allan variance for a laser frequency or beat frequency $\nu$, denoted by $\textrm{mod } \sigma^2_{\nu}$, is substituted in Eq.~\ref{eq:tch_mADEV} (taking the square on each side) by the \textit{single sided laser frequency noise power spectral density} $S_{\nu}(f)$ (as a function of the Fourier frequency $f$), the resultant equations still hold. Thus, based on the three-cornered-hat measurements, the noise power spectral density of an individual ultra-stable laser can be extracted in a manner similar to Eq.~\ref{eq:tch_mADEV}. Using this method, we determine the single sided frequency noise power spectral density of Laser1, denoted as $S_{\nu_{_\mathrm{Si1}}}(f)$.

	Frequency noise of lasers leads to a broadening of their line shapes. Based on the experimentally determined $S_{\nu_{_\mathrm{Si1}}}(f)$, we can compute a full width at half-maximum (FWHM) linewidth for Laser1. Here, although an exact relation between the frequency noise power spectral density and a FWHM linewidth exists~\cite{Elliott82pra}, an analytic form of this relation does not exist in the presence of a $1/f$ frequency noise. It is pointed out that the observed linewidth depends on the length of a measurement period~\cite{Domenico10appop}. Thus, by accounting for a finite measurement time, we can perform numerical computations to determine a minimum observable FWHM linewidth for a given set of measured $S_{\nu_{_\mathrm{Si1}}}(f)$.

	\begin{figure*}
		\centering
		\includegraphics[scale=0.56]{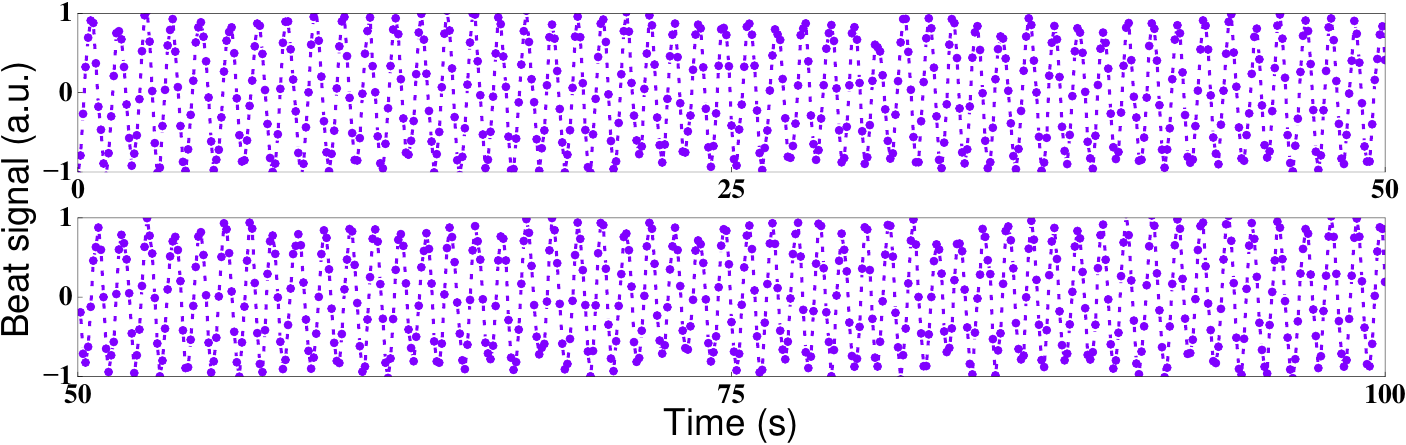}
		\caption{A sample data set for 100-s measurement (raw data) of the beat signal between Laser1 and Laser3 (mixed down to near d.c.). 
		}
		\label{fig:sm_sampleBeat100s}
	\end{figure*}

	Specifically, we follow a systematic methodology for laser linewidth determination given in Ref.~\cite{Bishof13prl}. Here, the power distribution of Laser1's electric field , $S_{E,\mathrm{Si1}}(\delta\nu)$, is given by~\cite{Bishof13prl} 
	\begin{small}
		\begin{eqnarray}\label{eq:SeFHWM}
			S_{E,{\mathrm{Si1}}}(\delta\nu) & = & 4E_0^2\int^{T_0}_0\left(1 - \frac{|\tau|}{T_0}\right)\cos\left(2\pi\delta\nu \tau\right) \nonumber\\
			& \times & \exp\left[-2\int^{\infty}_{1/T_0}S_{\nu_{_{\mathrm{Si1}}}}(f) \frac{\sin^2(\pi f \tau)}{f^2}\mathrm{d}f\right]\mathrm{d}\tau, \nonumber\\
		\end{eqnarray}
	\end{small}
	where $\delta\nu$ is the frequency deviation from Laser1's center frequency, $E_0$ is the amplitude of the electric field, $T_0$ is the measurement time over which the laser is observed. We compute and plot the $S_{E,\mathrm{Si1}}(\delta\nu)$ for different measurement time $T_0$, and determine the FWHM linewidth (based on the first half-maximum point closest to $\delta\nu = 0$) at each $T_0$. Finally, for $T_0$ up to 100~s, we determine the $T_0$ value under which the minimum observable linewidth of Laser1 is obtained. In this work, we find that the minimum observable linewidth is reached at $T_0 = 100$~s. Because 0.01~Hz is the lower limit we choose in this work for studying the noise power spectral measurement, we do not seek to investigate measurement time longer than 100~s and use the result under $T_0 = 100$~s as a conservative estimation of the ultra-narrow Laser1 linewidth. With $T_0 = 100$~s, we illustrate the above linewidth determination process by showing one individual sample set of noise power spectral density (see Fig.~\ref{fig:sm_compFWHMlinewidth}a) and the correspondingly computed FWHM linewidth of 9.4~mHz (see Fig.~\ref{fig:sm_compFWHMlinewidth}b). The histogram shown in Fig.~4c of the main text is obtained by analyzing multiple data sets (140 data sets that are 100~s long), and the line profile in the inset of Fig.~4c corresponds to the median value among the 140 FWHM linewidths.

	\begin{figure}
		\centering
		\includegraphics[scale=0.7]{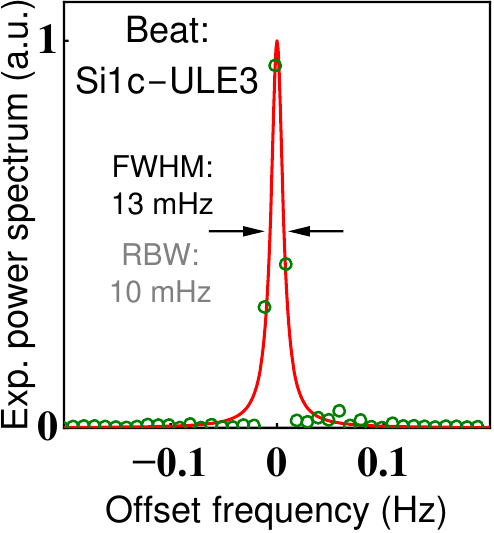}
		\caption{Power spectrum for the 100-s sample measurement of the Si1-ULE3 beat signal shown in Fig.~\ref{fig:sm_sampleBeat100s}. The power spectrum is obtained via fast Fourier transform with Hanning window. Here, FWHM denotes the full width at half-maximum, and RBW denotes the resolution bandwidth. In contrast to Fig.~4e that is the result of averaging twelve data sets, here, for demonstrating the quality of our data, we choose to show the result of one single data set with one of the narrowest linewidths.
		}
		\label{fig:sm_sampleBeatFFT}
	\end{figure}
	
	\subsection{Measurement and analysis of optical beat between two ultra-stable lasers}
	
	We measure the optical beat signal between Laser1 (Si1 cavity) and Laser3 (ULE3 cavity), where the Si1 cavity has been optimized to the Si1c condition with improvements (1) to (6) all implemented. The beat signal is mixed down to near d.c., and the corresponding time trace is subsequently recorded by a digital oscilloscope (Liquid Instruments, Moku:Pro). Here, to mitigate the influence of a linear frequency drift rate of the optical beat without implementing sophisticated drift-compensating feed-forward circuitry, we performed measurements in an approach based on the qualitative observation that the drift rates of optical beats vary relatively slowly with time (as showcased in Fig.~\ref{fig:sm_DriftRates}). In this approach, we first watched the Laser1-Laser3 beat signal on a K+K frequency counter until it showed a sufficiently small drift rate over several tens of seconds, and then promptly connect the beat signal (mixed down to near d.c.) to the digital oscilloscope to measure the time trace for 100 seconds. Likewise, we repeatedly waited for the appearance of near-zero-drift periods and accordingly measured time traces that were 100 seconds long. In one case, parts of a 100-s measurement (raw data) have been shown in Fig.~4d of the main text. In another case, a complete 100-s measurement (raw data) is shown in Fig.~\ref{fig:sm_sampleBeat100s}.
	
	To experimentally determine the narrowness of the  optical beat, we measure twelve  100-second-long data sets of such beat measurements. The raw data of each set is analyzed by fast Fourier transform (FFT) with Hanning window to extract the corresponding power spectrum. Each power spectrum is fitted with a Lorentzian profile, properly normalized, and then horizontally shifted to a position where the center of the corresponding individual fit sits at zero offset frequency. 
	Fig.~\ref{fig:sm_sampleBeatFFT} shows an example of such power spectrum based on the sample 100-s beat measurement in Fig.~\ref{fig:sm_sampleBeat100s}, exhibiting a fitted narrow linewidth of about 13~mHz for this individual data set, demonstrating the data quality for these measurements.
	Twelve  horizontally shifted power spectra are subsequently averaged (with a binsize of about 8.3~mHz) to yield the experimental result shown in Fig.~4e of the main text. The final averaged power spectrum is fitted by a Lorentzian profile to extract an ultra-narrow Si1-ULE3 beat linewidth of about 21(1)~mHz.
	Because our approach of waiting for a near-zero drift rate may not eliminate all influence of a linear frequency drift, the result of 21(1)~mHz can be viewed as an upper limit and a conservative experimental investigation for the intrinsic beat linewidth that would have been measured in the presence of an ideal linear frequency drift compensator.

	\section{Numerical simulations relevant to the horizontal fastening of the silicon cavity}

	\subsection{Effect of horizontal displacement of the silicon cavity}
	
	For room-temperature ULE cavities installed in a relatively quiet vacuum chamber with vibrational perturbation only from seismic sources, it is a standard method to discuss the vibration-induced frequency noise based on studies of vibrational acceleration and vibrational sensitivity. The underlying physical picture is that the vibrational noise is insufficient to cause a displacement of the cavity. However, in the case of few-kelvin cryogenic silicon cavities installed in a closed-cycle cryostat sample chamber, it is not warranted that the vibration stays below the limit of static friction. Rather, the cryostat vibration has a non-negligible probability to cause finite displacements, which will be particularly influential in the horizontal direction where the laser beam has a characteristic length scale of hundreds of micrometers or smaller. 
	
\begin{figure*}
		\centering
		\includegraphics[scale=0.5]{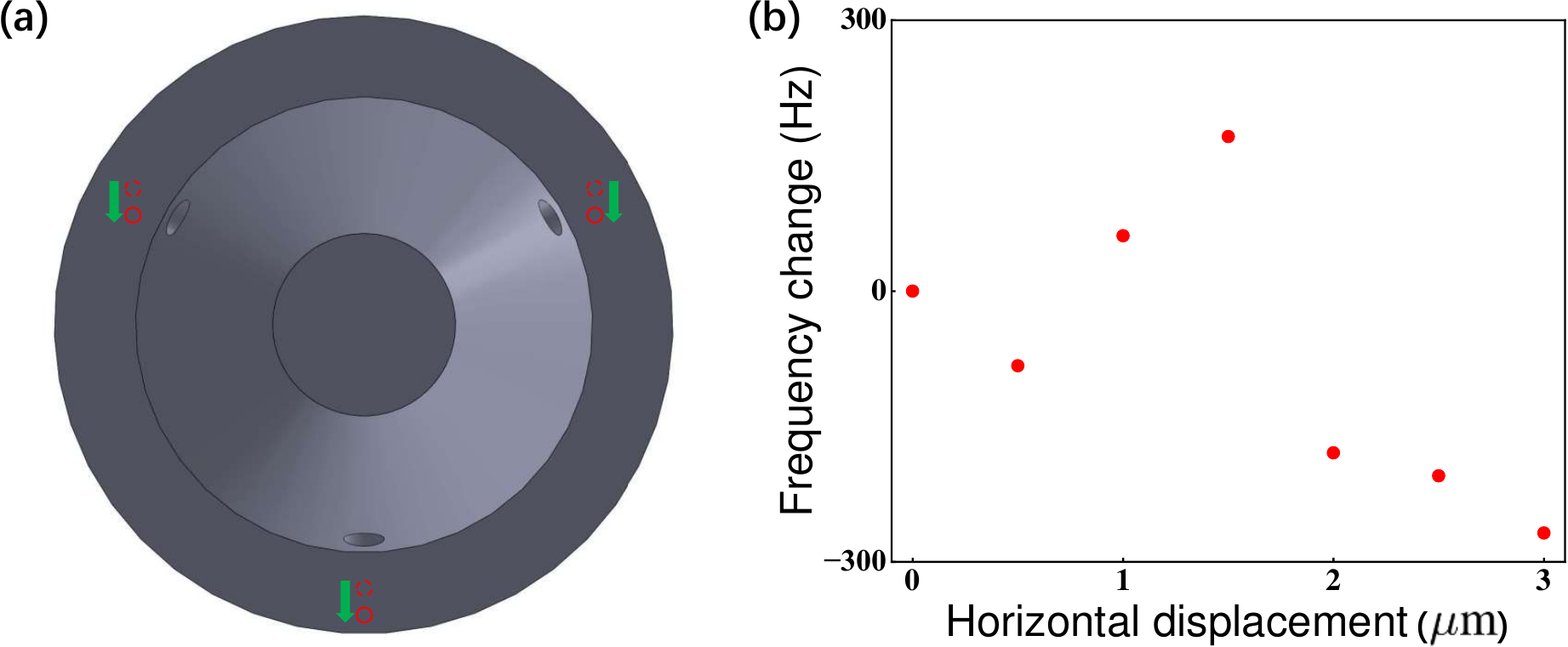}
		\caption{Effect of horizontal displacement of the silicon cavity (without horizontal PEEK fastening structure). (a) When the cavity has a horizontal displacement, the fixed three-point vertical supporting structure will have a relative displacement with respect to the cavity, as marked by the green arrows. (b) Change of the resonance frequency of the Si1 cavity as a function of the horizontal displacement value.
		}
		\label{fig:sm_horDisplacementEffect}
	\end{figure*}

	Taking this physical picture into account, we perform finite-element numerical simulations to quantify this effect. As illustrated in Fig.~\ref{fig:sm_horDisplacementEffect}a, when the cavity has a horizontal displacement and the vertical supporting structure is fixed, the three supporting points will have a relative displacement with respect to the cavity. We observe that, for displacements of $0.5\sim 3$~$\mu$m, the displacement of supporting points leads to resonance frequency change on the hundred-hertz level for the silicon cavity. Since the typical cryostat vibration corresponds to a displacement scale around the 100-nm level, such relative displacement will likely have a non-negligible effect on the silicon cavity's frequency instability, thus motivating the implementation of horizontal fastening of the silicon cavity in improvement (5). Here, we also emphasize that the actual \textit{in situ} cryostat vibration level at the position of the silicon cavity has not been well calibrated, and the analysis in this section is primarily meant to provide a qualitative motivation for improvement (5). The actual effect of cryostat vibration, without or with the implementation of improvement (5), is best illustrated in three-cornered-hat measurement of the Laser1 frequency instability, as shown in Fig.~3 and 4 of the main text, as well as Fig.~\ref{fig:sm_residualCryoVib} in this Supplementary material.

	\subsection{Effective CTE of the silicon cavity in the presence of improvement (5)}
	While we observe, for Si1b, vastly suppressed cryostat-vibration-induced frequency noise (shown in Fig.~3c and 3d of the main text) as a result of the implementation of improvement (5), it would be beneficial to study the effect of improvement (5) on the \textbf{effective} CTE for the silicon cavity installed in the horizontal fastening structure, such that the corresponding influence on the  long-term laser frequency stability can be evaluated.
	
	In a qualitative physical picture, at few-kelvin temperatures, single-crystalline silicon, stainless steel, and PEEK all have positive CTEs. When the system temperature decreases, the PEEK fastening structure shrinks, which will compress the center ring of the silicon cavity in the horizontal direction, and thus will lead to a correspondingly increase of the silicon cavity length along its optical axis in the vertical direction. Such an increase in the cavity length varies in a direction contrary to the cavity length decrease dictated by the CTE of single-crystalline silicon. In other words, the existence of horizontal PEEK fastening structure gives rise to a \textit{reduction in the effective CTE of the installed silicon cavity} with respect to a free silicon cavity under the same temperature.

	\begin{figure}
		\centering
		\includegraphics[scale=0.4]{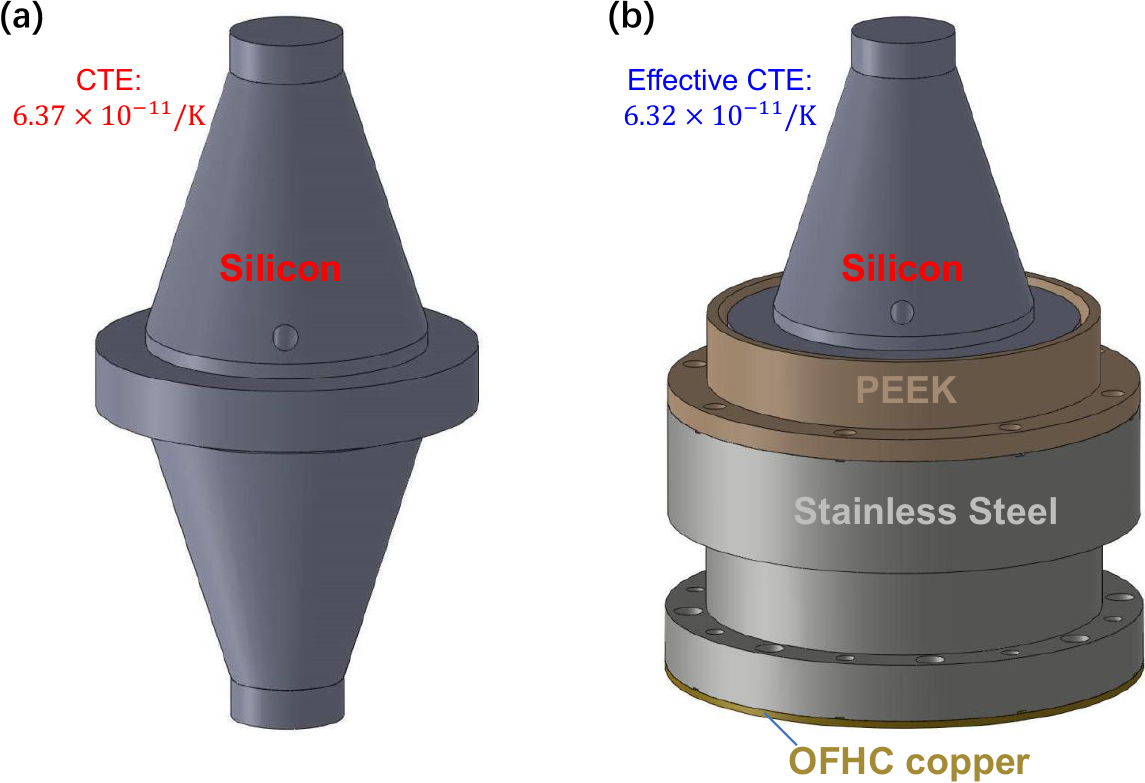}
		\caption{Effective coefficient of thermal expansion for a silicon cavity. (a) Effective CTE for a free silicon cavity. (b) Effective CTE for a silicon cavity installed in the whole vertically supporting and horizontally fastening structure.
		}
		\label{fig:sm_effectiveCTE}
	\end{figure}

	To complement this picture with quantitative evidences, we perform finite-element analysis with COMSOL numerical simulation software (shown in Fig.~\ref{fig:sm_effectiveCTE}). 
	\begin{itemize}
		\item We first simulate a free silicon cavity with a temperature change from 5~K to 6~K, which leads to a fractional frequency change of $6.37\times 10^{-11}$, corresponding to an average CTE of $6.37\times 10^{-11}$/K; see Fig.~\ref{fig:sm_effectiveCTE}a.
		\item Then, we simulate the actual whole assembly including the silicon cavity, the PEEK fastening structure (with typical CTE of $5\times 10^{-7}$/K under 5-K temperature~\cite{Wang25ijhe}), the stainless steel (with typical CTE not exceeding $2.5 \times 10^{-7}$/K under 5-K temperature~\cite{Ventura14SpringerBook}) supporting structure at which the PEEK structure is fixed, and a base made of oxygen-free high-conductivity copper (OFHC). With this whole assembly in the model and a boundary condition that fixes the bottom of the OFHC base, we observe a fractional frequency change of $6.32\times 10^{-11}$, corresponding to an average  \textbf{effective} CTE of $6.32 \times 10^{-11}$/K for the installed silicon cavity; see Fig.~\ref{fig:sm_effectiveCTE}b. 
		\item We also perform simulation for this assembly under a different boundary condition that allows the bottom of the stainless steel structure to freely expand/shrink, and observe for the installed silicon cavity an average effective CTE that is $9\%$ smaller than the CTE of a free silicon cavity.
	\end{itemize}

	Overall, considering that the stainless steel structure is fixed on a large base made of OFHC and that the OFHC CTE is at the $10^{-9}$/K level (two orders of magnitude smaller than that for stainless steel), we judge that the actual situation is much closer to the fixed-bottom boundary condition under which an effective CTE of $6.32\times 10^{-11}$/K is observed. Thus, we conclude that the effect of the horizontal fastening structure is a small decrease of the effective CTE of the silicon cavity, which is on the $10^{-13}$/K level (at most $10^{-12}$/K level). For the current operating temperature or future operating temperatures down to 3~K (where the free silicon CTE is about $1\times 10^{-11}$/K), such an effect of horizontal fastening is fairly small and actually beneficial because it decreases the effective CTE of the installed silicon cavity, which slightly reduces the long-term frequency instability due to cavity temperature fluctuation.

	Experimentally, with improvement (5) implemented, we perform a step-response measurement in a manner similar to Fig.~1d of the main text, and observe a total frequency change that agrees, to within $4\%$, with the prediction based on Fig.~1d for which improvement (5) has not been applied. This observation provides evidence that the effective CTE property of the Si1 cavity (installed in the assembly) indeed remains dominated by single-crystalline silicon material, which is consistent with the above physical picture based on numerical simulations.

	Here, we again note that \textbf{the key idea of the fastening structure is to use as little PEEK material in ``pressing the silicon center ring''  as possible such that not too much force is exerted onto the cavity}, which in this work is fulfilled by three small PEEK cylinders as parts of the monolithic PEEK structure to conservatively ensure sufficient overall rigidity. In the future, based on more advanced cryostat with suppressed cryostat vibration, the diameters of these three small cylinders can be further reduced, which can lead to even smaller reduction of the effective CTE of the installed silicon cavity, making the horizontal fastening scheme applicable to systems with even lower temperatures.

	\section{Experimental estimation of frequency noise induced by the residual cryostat vibration}
	With improvements (1) to (5) implemented, we observe that in Fig.~3d of the main text that, at Fourier frequencies below 20~Hz, the frequency noise power spectral density of Si1b reaches very similar level as that of Si1-CQQM. Here, in the Si1b configuration, the cryostat is operating at about 4.9~K. By contrast, in the Si1-CQQM configuration, the cryostat vibration is suspended after the system temperature reaches about 3.3~K, while other technical noise sources like seismic noise and opto-electrical noise stay the same as those in Si1b. The Brownian thermal noise floor in Si1b (about $3.3\times 10^{-17}$) is slightly larger than that in Si1-CQQM (about $2.7 \times 10^{-17}$), but is sufficiently close to the latter.
	
\begin{figure*}
		\centering
		\includegraphics[scale=0.55]{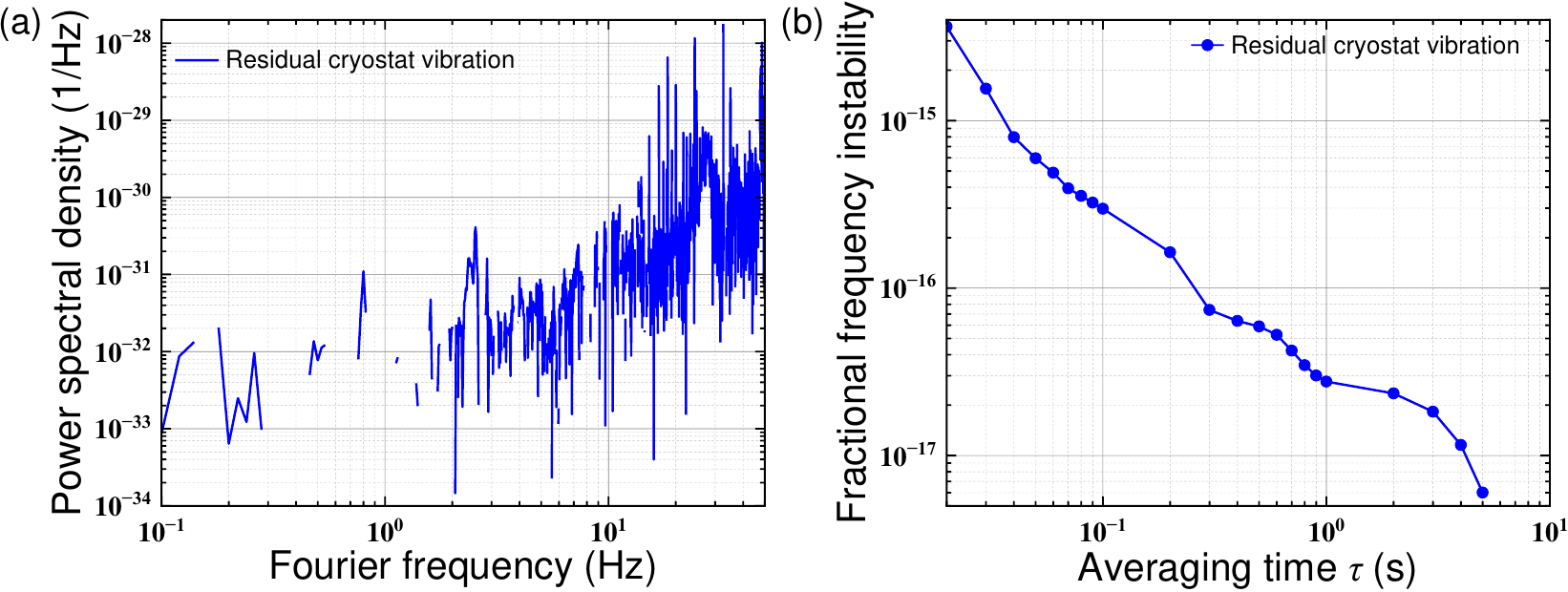}
		\caption{Effect of residual cryostat vibration on the Si1 cavity frequency instability. (a) The differential frequency noise power spectral density of Si1b versus Si1-CQQM based on the measurements shown in Fig.~3d of the main text. As a conservative estimation, we demand the differential power spectral density to be non-negative and set the differential power spectral density to be zero (thus not shown in panel (a)) when the Si1b power spectral density sits below that of Si1-CQQM. (b) Modified Allan deviation, computed based on the differential power spectral density in panel (a) and Eq.~\ref{eq:resVib}.
		}
		\label{fig:sm_residualCryoVib}
	\end{figure*}

	While this observation qualitatively shows that we have vastly suppressed the effect of cryostat vibration, we can further quantify the influence of residual cryostat vibration as follows. As shown in Fig.~\ref{fig:sm_residualCryoVib}a, we compute the differential frequency noise power spectral density, $S_{\mathrm{diff}}(f) = S_{\textrm{Si1b}}(f) - S_{\textrm{Si1-CQQM}}(f)$, between the Si1b and Si1-CQQM configurations, where $S_{\textrm{Si1b}}(f)$ and $S_{\textrm{Si1-CQQM}}(f)$ are the frequency noise power spectral densities for Si1b and Si1-CQQM respectively. As a conservative estimation, we demand $S_{\mathrm{diff}}(f)$ to be non-negative and set it to be zero when the Si1b power spectral density sits below that of Si1-CQQM. As another conservative estimation, we have also ignored the slight difference in the Brownian thermal noise of the two configurations. We then compute the corresponding modified Allan variance ($\textrm{mod }\sigma_{\textrm{res.cryo.vib.}}^2$) and modified Allan deviation ($\textrm{mod }\sigma_{\textrm{res.cryo.vib.}}$), of the residual cryostat vibration by numerically integrating $S_{\mathrm{diff}}(f)$ according to the following relation:
	\begin{eqnarray}\label{eq:resVib}
		&  & \textrm{mod }\sigma_{\textrm{res.cryo.vib.}}^2(\tau)   \nonumber\\
		& = &  2\int^{\infty}_0 S_{\mathrm{diff}}(f)\frac{\sin^6(\pi f \tau)}{(\pi f \tau^2/\tau_0)^2\sin^2(\pi f \tau_0)} \mathrm{d}f, \nonumber\\
	\end{eqnarray}
	where $\tau$ is the averaging time, $f$ is the Fourier frequency, $\tau_0 = 0.01$~s is the temporal sampling resolution. As shown in Fig.~\ref{fig:sm_residualCryoVib}b, the residual cryostat vibration gives rise to a modified Allan deviation that reduces to a level below $2\times 10^{-17}$ for $\tau \ge 3$~s, which is an improvement by at least an order of magnitude over the cryostat vibration effect that can be derived from the contrast between Si1a and Si1-CQQM (see Fig.~3a of the main text).

	\section{Determination and control of individual noise sources}

	\begin{figure}[t]
		\includegraphics[width = 7.5cm]{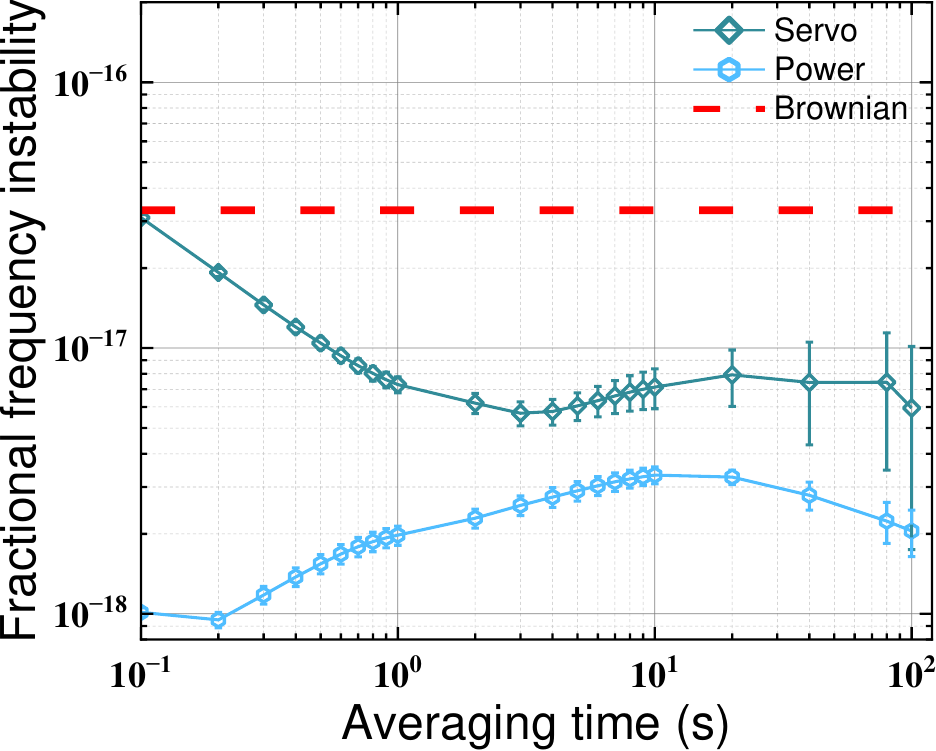}
		\caption{\label{fig:sm_indivinoise} Frequency instabilities of the Si1 cavity induced by several noise sources. Measured results are shown for  PDH servo noise (bluish green diamonds) and power fluctuation noise (cyan hexagons). Projected instabilities are computed for the Brownian thermal noise floor of Si1 cavity at 4.9~K (red dashed line). Error bars represent $1\sigma$ statistical uncertainties.			
		}
	\end{figure}

	In this section, we describe the methods to determine, estimate, and control various individual noise sources that affect the frequency stability of the Si1 cavity. The primary motivation of these individual measurements and estimations is to promote better understanding on the results of the three-cornered-hat measurements shown in Fig.~4a of the main text. \\

	\subsection{The fundamental Thermal noise floor}
	Thermal noise is the fundamental noise floor of a given ultra-stable cavity. 
	The single sided power spectral density of the thermal noise displacement ($G_{x,\mathrm{total}}$) has the unit of  $\mathrm{m}^2$/Hz and consists of  Brownian motion of the high-reflectivity mirror coating ($G_{x,\mathrm{c}}$), Brownian motion of the spacer ($G_{x,\mathrm{spacer}}$), Brownian motion of the mirror substrate ($G_{x,\mathrm{sub}}$), as well as the substrate thermo-elastic noise ($G_{x,\mathrm{TE}}$) and coating thermal-optic noise ($G_{x,\mathrm{TO}}$)~\cite{Cole13nph}:
	\begin{eqnarray}
		G_{x,\mathrm{total}}(f) & = & G_{x,\mathrm{c}}(f) + G_{x,\mathrm{spacer}}(f) + G_{x,\mathrm{sub}}(f) \nonumber\\
		& + &  G_{x,\mathrm{TE}}(f) + G_{x,\mathrm{TO}}(f) 
	\end{eqnarray}
	The three Brownian thermal noise sources are given by the following expressions~\cite{Cole13nph}:
	\begin{eqnarray}
		G_{x,\mathrm{spacer}}(f) & = & \frac{2k_{\mathrm{B}}T}{\pi^2 f}\frac{L}{(R^2 - r^2)Y_{\mathrm{spacer}}}\phi_{\mathrm{spacer}}, \nonumber\\
		G_{x,\mathrm{sub}}(f) & = & \frac{4k_{\mathrm{B}}T}{\pi^{3/2}f}\frac{1-\sigma^2_{\mathrm{sub}}}{w_{\mathrm{m}}Y_{\mathrm{sub}}}\phi_{\mathrm{sub}}, \nonumber\\
		G_{x,\mathrm{c}}(f) & = & \frac{4k_{\mathrm{B}}T}{\pi^2 f}\frac{1-\sigma^2_{\mathrm{sub}}}{w_\mathrm{m}Y_{\mathrm{sub}}}\frac{D}{w_{\mathrm{m}}}a_{x,\mathrm{c}}(b_{x,\mathrm{c}}+c_{x,\mathrm{c}}), \nonumber
	\end{eqnarray}
	where $L$ is the cavity length, $w_m$ the $1/e^2$ radius of the incident laser beam, $R$ the average outer radial size of the spacer, $r$ the radius of the spacer's center bore for laser propagation, and $a_{x,\mathrm{c}}$, $b_{x,\mathrm{c}}$, $c_{x,\mathrm{c}}$ are given by
	\begin{eqnarray}
		a_{x,\mathrm{c}} & = & \frac{\phi_{\mathrm{c}}}{Y_{\mathrm{sub}}Y_{\mathrm{c}}(1-\sigma^2_{\mathrm{c}})(1-\sigma^2_{\mathrm{sub}})} \nonumber\\
		b_{x,\mathrm{c}} & = & Y^2_{\mathrm{c}}(1+\sigma_{\mathrm{sub}})^2(1-2\sigma_{\mathrm{sub}})^2\nonumber\\
		c_{x,\mathrm{c}} & = & Y^2_{\mathrm{sub}} (1+\sigma_{\mathrm{c}})^2 (1-2\sigma_{\mathrm{c}}). \nonumber
	\end{eqnarray}
	The other two thermal noise sources are given by the following expressions~\cite{Cole13nph}:
	\begin{eqnarray}
		G_{x,\mathrm{TE}}(f) & = & \frac{8}{\sqrt{\pi}}(1+\sigma_{\mathrm{sub}})^2\alpha^2_{\mathrm{sub}}\frac{k_{\mathrm{B}}T^2w_{\mathrm{m}}}{\kappa_{\mathrm{sub}}}\times\int^{+\infty}_0\mathrm{d}u \nonumber\\
		& &  \int^{+\infty}_{-\infty}\mathrm{d}v \frac{\sqrt{2/\pi^3}u^3e^{-u^2/2}}{(u^2 + v^2)\left[(u^2+v^2)^2 + \Omega^2_{\mathrm{TE}}(f)\right]}, \nonumber\\
		G_{x,\mathrm{TO}}(f) & = & \frac{4k_{\mathrm{B}}T^2}{\sqrt{\pi}w_{\mathrm{m}}\kappa_{\mathrm{sub}}} d_{\mathrm{TO}}^2 \textrm{ , when }\Omega_{\mathrm{TE}}(f) << 1,\nonumber
	\end{eqnarray}
	where
	\begin{eqnarray}
		\Omega_{\mathrm{TE}}(f) & = & w^2_{\mathrm{m}}C_{\mathrm{sub}}\pi f/\kappa_{\mathrm{sub}}, \nonumber
	\end{eqnarray}
	and 
	\begin{eqnarray}
		d_{\mathrm{TO}} & = & \bar{\alpha}_{\mathrm{c}}D - \bar{\beta}_{\mathrm{c}}\lambda - \alpha_{\mathrm{sub}}DC_{\mathrm{c}}/C_{\mathrm{sub}} \nonumber.
	\end{eqnarray}
	
\begin{figure}[htbp]
		\includegraphics[width = 7.6cm]{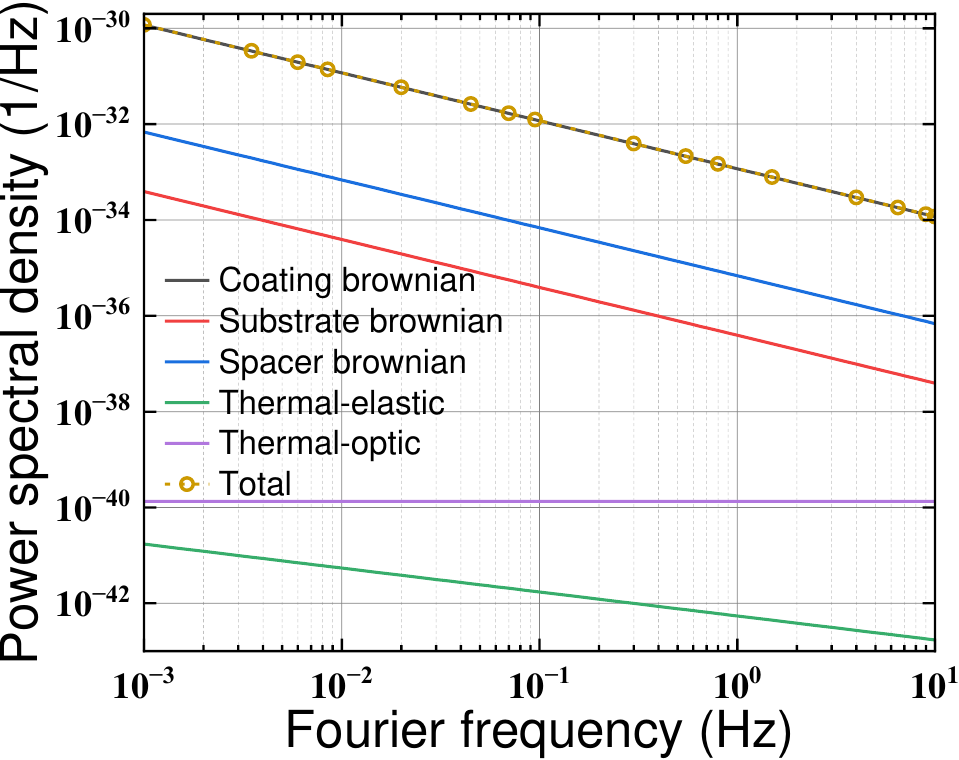}
		\caption{\label{fig:sm_ThermalNoiseSources}  Computed power spectral density for thermal noise of the Si1 cavity at 4.9~K, with contributions from five sources. Here, the total thermal noise (yellowish brown empty circles with dashed line) almost overlaps with the dominant contribution from the coating Brownian noise (dark gray line). 
		}
	\end{figure}

\begin{figure}[htbp]
		\includegraphics[width = 7.6cm]{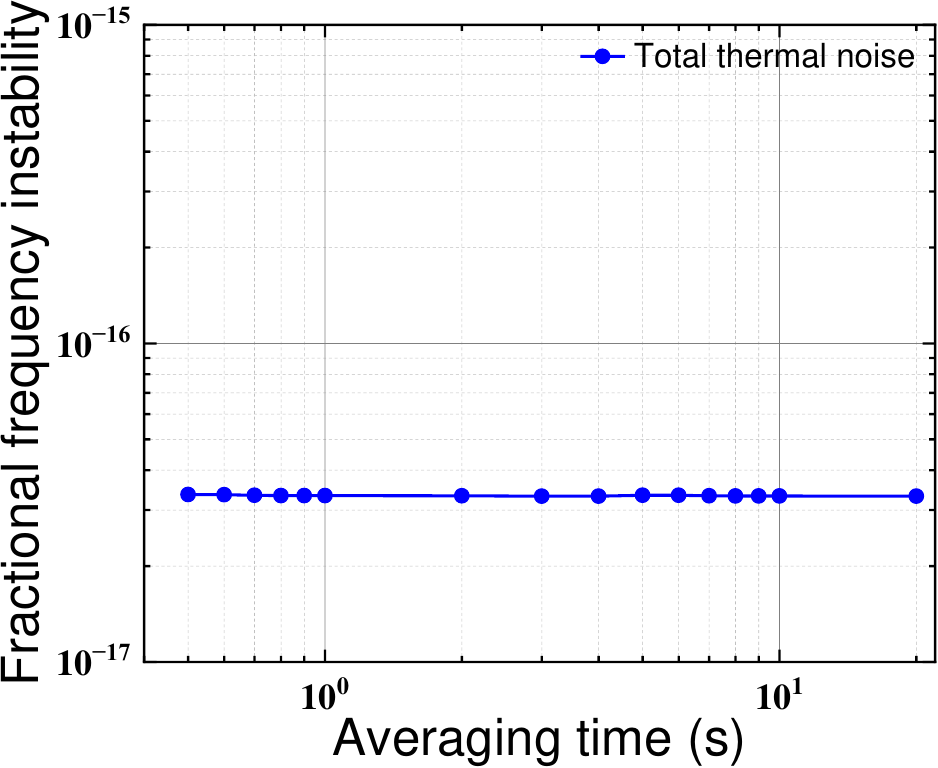}
		\caption{\label{fig:sm_thermal_mADEV}  Computed modified Allan deviation for the total thermal noise of Si1 cavity at 4.9~K, which is dominated by the Brownian thermal noise of the coating. 
		}
	\end{figure}

	The frequency noise power spectral density can be derived from the displacement noise power spectral density as
	\begin{eqnarray}
		S_{\nu}(f) & = & G_{x,\mathrm{total}}\frac{\nu^2}{L^2},
	\end{eqnarray}
	where $L$ is the cavity length and $\nu$ is the cavity resonance frequency. In Table~\ref{table:thermalNoiseParameters}, we list out the relevant parameters for computing the thermal noise floor of the Si1 cavity for 1397~nm and $T = 4.9$~K, which is dominated by the Brownian thermal noise (see Fig.~\ref{fig:sm_ThermalNoiseSources}), with a computed modified Allan deviation of $3.3\times 10^{-17}$ (see Fig.~\ref{fig:sm_thermal_mADEV}). The Brownian thermal noise at other temperatures in the few-kelvin regime can be obtained according to its $\sqrt{T}$ dependence. With a set of parameters similarly chosen for 1542~nm, our computation has also \textit{reproduced} the $6.5\times 10^{-17}$ Brownian thermal noise floor for the 4-K silicon cavity reported in Refs.~\cite{Zhang17prl, Robinson19optica}.\\

	\subsection{Technical noise} 
	\textit{Residual amplitude modulation (RAM).---}We determine the RAM for Si1 cavity by measuring voltage fluctuation of the off-resonance, out-of-loop PDH error signal and multiplying it by the voltage-to-frequency PDH discriminator slope. Based on a wedged facet of the free-space resonant electro-optic phase modulator (Qubig, PM7-SWIR), unwanted polarization component can be spatially separated from the main beam. With improvements (1) to (6) all implemented, the measured RAM for Si1 cavity is shown as gray dashed line in Fig.~4a in the main text.\\ 
	
	\begin{table}
		\center
		\caption{Symbols and parameters used in the thermal noise computations for the Si1 cavity. Here, ``c'' denotes the dielectric coating, ``sub'' and ``spacer'' in the subscript denote the single-crystalline mirror substrate and cavity spacer. The values of these parameters are chosen based on a series of literatures~\cite{Robinson21ol, Kedar23thesis,Martin13thesis,Evans08prd, Ho72jpcrd, Desai86jpcrd}.}
		\label{table:thermalNoiseParameters}
		\begin{tabular}{lll} 
			\specialrule{1.0pt}{0pt}{0pt}
			Parameter & Description & Value \\
			\specialrule{0.5pt}{0pt}{0pt}
			$L$ & cavity length & 9.79~cm\\
			MG1 & Mirror1 Geometry & PL/PL\\
			MG2 & Mirror2 Geometry & PL/CC, R=1~m\\
			$T$ & Temperature  & 4.9~K\\
			$Y_{\mathrm{sub}/\mathrm{spacer}}$ & Young's modulus  &  187.5~GPa \\
			& for the substrate/spacer & \\
			$\sigma_{\mathrm{sub}/\mathrm{spacer}}$& Poisson's ratio & 0.23\\
			& for the substrate/spacer & \\
			$\phi_{\mathrm{sub}/\mathrm{spacer}}$& Loss angle & $1\times 10^{-7}$\\
			& for the substrate/spacer& \\
			$\alpha_{\mathrm{sub}/\mathrm{spacer}}$& Coef. of thermal expansion & $4.8\times 10^{-11}$/K \\
			& for the substrate/spacer& \\
			$\kappa_{\mathrm{sub}/\mathrm{spacer}}$& Thermal conductivity & 537~W/(m$\cdot$K)\\
			& for the substrate/spacer & \\
			$C_{\mathrm{sub}/\mathrm{spacer}}$ & Heat capacity per volume & 80.43~J/(K$\cdot \mathrm{m}^3$) \\
			& for the substrate/spacer & \\
			$Y_{\mathrm{c}}$ & Young's modulus for coating & 110~GPa \\
			$\sigma_{\mathrm{c}}$ & Poisson's ratio for coating & 0.2 \\
			$\phi_{\mathrm{c}}$ & Loss angle for coating & $7\times 10^{-4}$ \\
			$\bar{\alpha}_{\mathrm{c}}$ & Coef. of thermal expansion & $4.47\times 10^{-6}$/K \\
			& for coating & \\
			$\kappa_{\mathrm{c}}$ & Thermal conductivity  & 2.65~W/(m$\cdot$K)\\
			& for coating & \\
			$C_{\mathrm{c}}$ & Heat capacity per volume & $1.87\times 10^6$\\
			& for coating & ~~J/(K$\cdot \mathrm{m}^3$)\\
			$\bar{\beta}_{\mathrm{c}}$ & Effective thermorefractive &  $6.86\times 10^{-6}$/K\\
			& coef. for coating & \\
			$D$ & Thickness for coating & 8.3~$\mu$m\\
			\specialrule{1.0pt}{0pt}{0pt}
		\end{tabular}
	\end{table}
	
	\textit{PDH servo noise.---}We determine the PDH servo noise by measuring the voltage fluctuation in the in-loop PDH error signal  and multiplying it by the voltage-to-frequency discriminating slope (1.87~kHz/V). The measured PDH servo noise is shown as Bluish green diamonds in Fig.~\ref{fig:sm_indivinoise}.\\
	
	\textit{Optical power fluctuation.---}To control the frequency noise induced by optical power fluctuation of the beam entering Si1 cavity, we implement active feedback control of the optical power of this laser beam, and then determine this frequency instability using a similar method as that for the PDH servo effect. Here, the incident power is about 1.1~$\mu$W before the beam enters the bottom viewport of the sample vacuum chamber; the associated power fluctuation is measured via the d.c. voltage output of the PDH photo-detector, and is subsequently multiplied by the power-to-frequency conversion coefficient (about 2.6(1)~Hz/$\mu$W). As shown by cyan hexagons in Fig.~\ref{fig:sm_indivinoise}, the power fluctuation gives rise to frequency instabilities below $1\times 10^{-17}$ at shown averaging times.\\

	\textit{Seismic vibrational noise.---}Here, we use ``seismic vibrations'' to denote vibrations from the environment (excluding the vibration from the cryostat coolers). We estimate the frequency noise induced by seismic vibrations based on measurements of (1) vibrational noise on the breadboard where the sample chamber sits and (2) vibrational sensitivities along three spatial directions.
	\begin{itemize}
		\item For measuring the vibrational noise, we use a combination of an uniaxial accelerometer (Endevco, Model 86) and a triaxial seismometer (G\"uralp, 3T-120). The accelerometer has better sensitivity at shorter time scales and is applied for averaging times below 0.6~s; the seismometer has better sensitivity at longer time scales and is used for averaging times above 0.6~s. 
		\item For measuring vibrational sensitivities, we induce external modulation at a Fourier frequency of 6~Hz by a commercial modulation input module (Table Stable, AMB-2), and simultaneously measure (a) the vibrational noise amplitude at 6~Hz on the breadboard where the sample vacuum chamber sits and (b) the corresponding frequency modulation of the silicon laser (via three-cornered-hat measurements). The vibrational sensitivity along a certain spatial direction is extracted as a proportionality coefficient between the frequency modulation and the vibrational noise amplitude.
	\end{itemize} 
	\noindent After measuring the power spectral densities of vibrational noise in three spatial directions and calibrating the corresponding vibration sensitivities, we perform numerical integration to extract the modified Allan deviation~\cite{Hagemann13thesis}, $\sigma_{_\mathrm{V}}$, for the seismic-vibration-induced frequency instability, which is shown as a blue dashed line in Fig.~4a of the main text.\\

	\textit{Cavity temperature fluctuation.---}We compute a characteristic frequency instability induced by the Si1 cavity temperature fluctuation based on the Si1 operating temperature (about 4.9~K), knowledge of the silicon CTE (about $4.5\times 10^{-11}$/K; see Ref.~\cite{Kedar23thesis} and the main text), and the Si1 cavity temperature fluctuation. Here, the Si1 cavity temperature fluctuation is in turn computed based on measured temperature instability of the ``second active shield'' of the cryostat (where the temperature is actively controlled at the Si1 operating temperature by a feedback servo) and measured time constants for heat transfer between that shield and the Si1 cavity. The typical temperature instability of the second active shield is measured to reach 200~$\mu$K at 4 seconds. With improvements (1) to (6) all implemented, we measure a set of time constants of 4400~s, 330~s, 330~s for heat transfer between the second active shield and the silicon cavity. These time constants are slightly different from that extracted from Fig.~1d in the main text because improvement (5) is implemented here and absent in Fig.~1d of the main text. With the above performance for thermal control and passive damping, we compute characteristic silicon cavity temperature instabilities of $4\times$ and $8\times 10^{-8}$~K at averaging times of 2 and 4 seconds. The projected temperature-fluctuation-induced silicon laser frequency instability is shown as a brown dashed line in Fig.~4a of the main text, which stays below the Brownian thermal noise for averaging time of $\tau \le 38$~s. Our measurement (shown in the inset of Fig.~4a of the main text) at a lower platform temperature of 3.3~K further supports that the Si1c long-term frequency instability (shown in Fig.~4a of the main text) is dominated by the Si1 temperature fluctuations; see also Sec.~\ref{sm_sec:TCH3p3K}.\\

\begin{figure}[t]
		\includegraphics[width = 7.6cm]{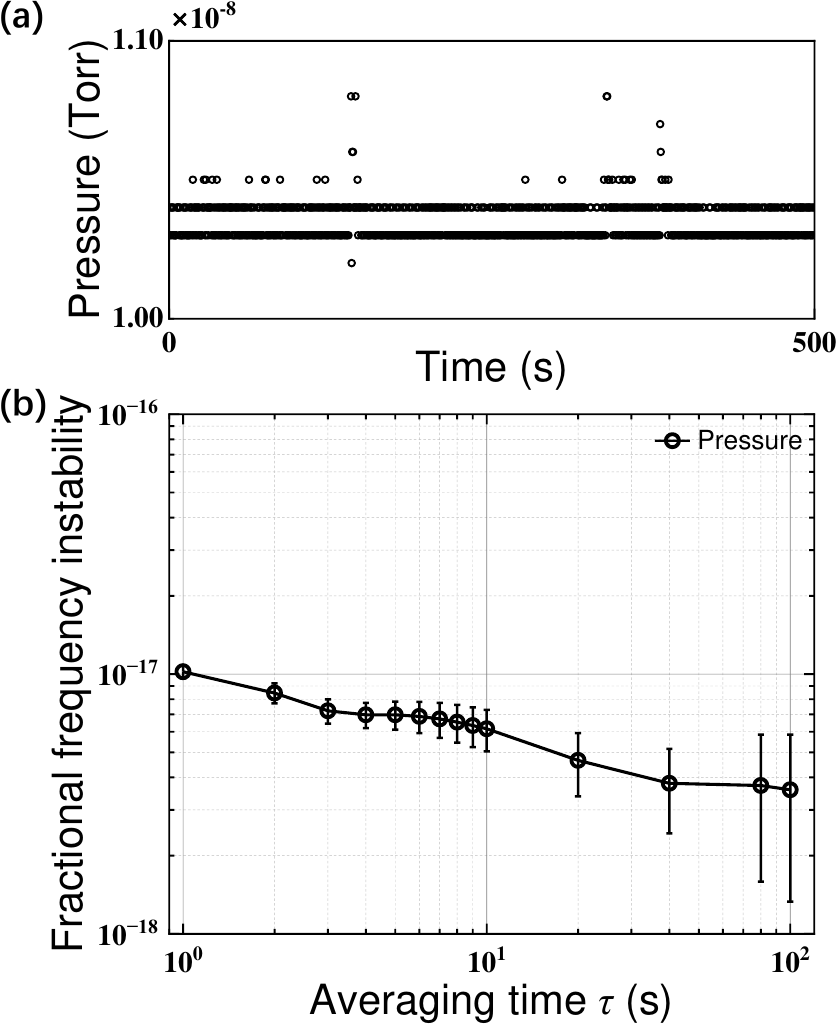}
		\caption{\label{fig:sm_pressure} Measurement of pressure fluctuation inside the ion pump connected to the cryostat sample chamber and estimation of the associated Si1 frequency instability, under a 4.9-K platform temperature. (a) Pressure fluctuation during a 500-second measurement based on readings of the ion pump permanently connected to the cryostat sample chamber. (b) Estimation of the pressure-fluctuation-induced fractional frequency instability, computed via multiplying the measured pressure fluctuation by a conversion coefficient of $3.7 \times 10^{-7}$/Torr used in Ref.~\cite{Kedar23thesis}.
			Error bars represent $1\sigma$ statistical uncertainties.
		}
	\end{figure}

	\textit{Pressure fluctuation.---}Because we do not have a vacuum gauge that sits immediately next to the silicon cavity, we use the reading of the ion pump 
	that is permanently connected to the cryostat sample chamber, as an estimation of the pressure at the place of the silicon cavity. The actual pressure at the cavity is likely lower due to the powerful cryogenic pumping at sub-5~K. Fig.~\ref{fig:sm_pressure}a shows a measurement of pressure fluctuation based on the ion pump reading. After multiplying the pressure fluctuation by a conversion coefficient of $3.7\times 10^{-7}$/Torr between the fractional frequency change and the pressure change~\cite{Kedar23thesis}, we compute the pressure-fluctuation-induced fractional frequency instability. As shown in Fig.~\ref{fig:sm_pressure}b, the typical fractional frequency instability stays around $1\times 10^{-17}$ and lower levels for averaging time in the range of 1 to 100~s, which is sufficiently small and thus neglected in this article.

	\section{Improved Si1 long-term frequency stability at a lower operating temperature}\label{sm_sec:TCH3p3K}
\begin{figure}[htbp]
		\includegraphics[width = 7.6cm]{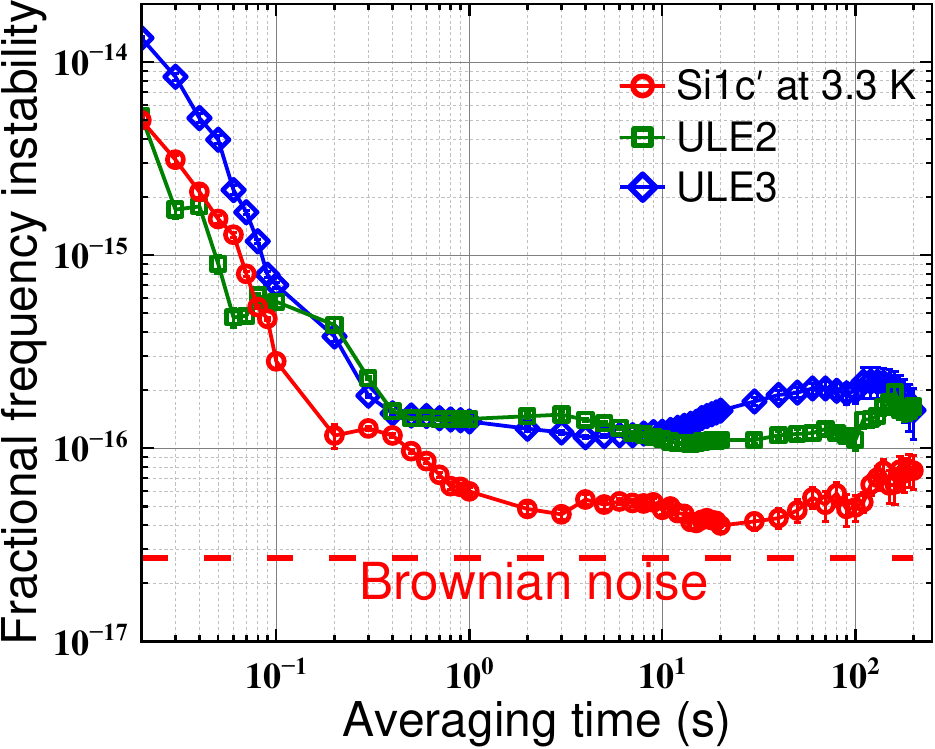}
		\caption{\label{fig:sm_Si1at3p3K} Modified Allan deviations based on three-cornered-hat measurement. Here, Si1c$'$ represents the Si1 cavity at a lower operating temperature (with platform temperature of 3.3~K), which is otherwise identical to the setup of Si1c. For a 3.3-K operating temperature, the Brownian thermal noise floor of the Si1 cavity reduces to about $2.7\times 10^{-17}$.	Error bars represent $1\sigma$ statistical uncertainties. The data for the Si1c$'$ frequency instability and the Brownian noise at 3.3~K in this figure are also plotted in the inset of Fig.~4a in the main text.
		}
	\end{figure}

	While the performance of the Si1 cavity well enters the $10^{-17}$ regime for averaging times up to 100~s, it is beneficial to investigate the primary reason why the Si1c frequency instability actually increases for averaging times exceeding 20~s. Based on the horizontal fastening of the Si1 cavity, the influence of cryostat cooler vibration has been vastly suppressed, making it possible to decrease the operating temperature again such that \textbf{the silicon CTE is further reduced}. We performed such a measurement by choosing the ``Medium'' cooling mode instead of the ``Slow'' cooling mode used for Fig.~4 of the main text. The resultant platform temperature decreases from 4.9~K to 3.3~K, which leads to a change of the Si1 Brownian thermal noise floor from $3.3 \times 10^{-17}$ to $2.7 \times 10^{-17}$, where we denote this lower-temperature operating condition as Si1c$'$ for the silicon cavity. Accordingly, we observe that the Si1c$'$ frequency instability remains fairly flat around $4\sim 5\times 10^{-17}$ for a wide range of averaging times of $2\sim 100$~s, which shows substantially better long-term instability than that of Si1c at 4.9~K; see Fig.~\ref{fig:sm_Si1at3p3K}. The Si1c$'$ frequency instability starts to increase with $\tau$ only after $\tau$ exceeds 100~s. The contrast between long-term frequency stabilities of Si1c$'$ at 3.3~K and Si1c at 4.9~K shows that the cavity temperature fluctuation under a finite silicon CTE remains the primary reason why the Si1c frequency instability increases at relatively long averaging times.

	To further reveal the present long-term performance of the silicon cavity under the  Si1c$'$ configuration, we extended the long-term frequency instability measurement to a range of $\tau \le 1000$~s, observing a frequency instability of $5.5(1.2)\times 10^{-16}$ at $\tau = 1000$~s, as shown in the \textit{inset} of Fig.~4a in the main text. Further optimizing the Si1 frequency stability at relatively long averaging times is a sufficiently important and challenging question that is outside the scope of this article, which will be studied in future researches.
	
\end{document}